\documentclass[journal=ancac3, manuscript=article]{achemso}

\usepackage[version=3]{mhchem} % Formula subscripts using \ce{}
\usepackage[english]{babel}
\usepackage[utf8]{inputenc} % proper recognition of "ü", "ö" and 
\usepackage{siunitx} % proper unit formatting
\usepackage{hyperref} % needed for \sfigref

\mciteErrorOnUnknownfalse

\newcommand{\tsub}[1]{\textsubscript{#1}}
\newcommand{\tcm}{\,cm\textsuperscript{-1}}
\newcommand{\lex}{$\lambda_{\text{exc}}\,$}

\newcommand*{\sfigref}[2][]{% use as: "... as seen in Figure~\sfigref[b]{fig:Fig1} ..."
  \hyperref[{#2}]{%
    \ref*{#2}%
    \ifx\\#1\\%
    \else
      \,#1)% formatting of first argument
    \fi
  }%
}

\AtBeginDocument{
\begin{center}
    \textbf{Preprint\\}
\end{center}

This document is the unedited Author’s version of a submitted work that was subsequently accepted for publication in ACS Nano, copyright \copyright~American Chemical Society after peer review.\\

To access the final edited and published work see:\\

Overbeck, Jan; Borin Barin, Gabriela; Daniels, Colin; Perrin, Mickael L.; Braun, Oliver; Sun, Qiang; Darawish, Rimah; Luca, Marta de; Wang, Xiao-Ye; Dumslaff, Tim; Narita, Akimitsu; Müllen, Klaus; Ruffieux, Pascal; Meunier, Vincent; Fasel, Roman; Calame, Michel\\

\textbf{A Universal Length-Dependent Vibrational Mode in Graphene Nanoribbons}

\textit{ACS Nano} \textbf{2019}, 13, 13083–13091

DOI: 10.1021/acsnano.9b05817\\

Access to the published version may require subscription.
When citing this work, please cite the original published paper.
\newpage
}

\author{Jan Overbeck} 
\affiliation{Empa, Swiss Federal Laboratories for Materials Science and Technology, 8600 Dübendorf, Switzerland}
\alsoaffiliation{University of Basel, Department of Physics, 4056 Basel, Switzerland}
\alsoaffiliation{University of Basel, Swiss Nanoscience Institute, 4056 Basel, Switzerland}
\altaffiliation{These authors contributed equally.}
\email{jan.overbeck@empa.ch}

\author{Gabriela Borin Barin} 
\affiliation{Empa, Swiss Federal Laboratories for Materials Science and Technology, 8600 Dübendorf, Switzerland}
\altaffiliation{These authors contributed equally.}

\author{Colin Daniels}
\affiliation{Rensselaer Polytechnic Institute, Department of Physics, Applied Physics, and Astronomy, Troy, New York 12180, United States}
\altaffiliation{These authors contributed equally.}

\author{Mickael Perrin}
\affiliation{Empa, Swiss Federal Laboratories for Materials Science and Technology, 8600 Dübendorf, Switzerland}

\author{Oliver Braun}
\affiliation{Empa, Swiss Federal Laboratories for Materials Science and Technology, 8600 Dübendorf, Switzerland}
\alsoaffiliation{University of Basel, Department of Physics, 4056 Basel, Switzerland}

\author{Qiang Sun}
\affiliation{Empa, Swiss Federal Laboratories for Materials Science and Technology, 8600 Dübendorf, Switzerland}

\author{Rimah Darawish}
\affiliation{Empa, Swiss Federal Laboratories for Materials Science and Technology, 8600 Dübendorf, Switzerland}
\alsoaffiliation{University of Bern, Department of Chemistry and Biochemistry, 3012 Bern, Switzerland}

\author{Marta De Luca} % Raman data analysis
\affiliation{University of Basel, Department of Physics, 4056 Basel, Switzerland}

\author{Xiao-Ye Wang} %xiaoye.wang@nankai.edu.cn

\affiliation{Max Planck Institute for Polymer Research, 55128 Mainz, Germany}
\altaffiliation{Current address: State Key Laboratory of Elemento-Organic Chemistry, College of Chemistry, Nankai University, Tianjin 300071, China}
\author{Tim Dumslaff}% synthesized
\affiliation{Max Planck Institute for Polymer Research, 55128 Mainz, Germany}

\author{Akimitsu Narita} % synthesized
\affiliation{Max Planck Institute for Polymer Research, 55128 Mainz, Germany}

\author{Klaus Müllen} % supervised
\affiliation{Max Planck Institute for Polymer Research, 55128 Mainz, Germany}
\alsoaffiliation{Institute of Physical Chemistry, Johannes Gutenberg-Universität Mainz, 55128 Mainz, Germany}

\author{Pascal Ruffieux}
\affiliation{Empa, Swiss Federal Laboratories for Materials Science and Technology, 8600 Dübendorf, Switzerland}

\author{Vincent Meunier}
\affiliation{Rensselaer Polytechnic Institute, Department of Physics, Applied Physics, and Astronomy, Troy, New York 12180, United States}
\email{meuniv@rpi.edu}

\author{Roman Fasel}
\affiliation{Empa, Swiss Federal Laboratories for Materials Science and Technology, 8600 Dübendorf, Switzerland}
\alsoaffiliation{University of Bern, Department of Chemistry and Biochemistry, 3012 Bern, Switzerland}
\email{roman.fasel@empa.ch}

\author{Michel Calame}
\affiliation{Empa, Swiss Federal Laboratories for Materials Science and Technology, 8600 Dübendorf, Switzerland}
\alsoaffiliation{University of Basel, Department of Physics, 4056 Basel, Switzerland}
\alsoaffiliation{University of Basel, Swiss Nanoscience Institute, 4056 Basel, Switzerland}
\email{michel.calame@empa.ch}

\title[LCM in GNRs]{A Universal Length-Dependent Vibrational Mode in Graphene Nanoribbons}


\begin{document}

%\begin{tocentry}

% Some journals require a graphical entry for the Table of Contents.
% This should be laid out ``print ready'' so that the sizing of the

% Inside the \texttt{tocentry} environment, the font used is Helvetica
% 8\,pt, as required by \emph{Journal of the American Chemical

% The surrounding frame is 9\,cm by 3.5\,cm, which is the maximum
% permitted for  \emph{Journal of the American Chemical Society}
% graphical table of content entries. The box will not resize if the
% content is too big: instead it will overflow the edge of the box.

% This box and the associated title will always be printed on a
% separate page at the end of the document.

%\includegraphics[height=36mm]{Figures/Graphical-Abstract.pdf}

%\end{tocentry}

\newpage
\begin{abstract}

Graphene nanoribbons (GNRs) have attracted considerable interest as their atomically tunable structure makes them promising candidates for future electronic devices. However, obtaining detailed information about the length of GNRs has been challenging and typically relies on low-temperature scanning tunneling microscopy. Such methods are ill-suited for practical device application and characterization. In contrast, Raman spectroscopy is a sensitive method for the characterization of GNRs, in particular for investigating their width and structure. 
Here, we report on a length-dependent, Raman active low-energy vibrational mode that is present in atomically precise, bottom-up synthesized armchair graphene nanoribbons (AGNRs). 
Our Raman study demonstrates that this mode is present in all families of AGNRs and provides information on their length. Our spectroscopic findings are corroborated by scanning tunneling microscopy images and supported by first-principles calculations that allow us to attribute this mode to a longitudinal acoustic phonon. 
Finally, we show that this mode is a sensitive probe for the overall structural integrity of the ribbons and their interaction with technologically relevant substrates.\\
%\textbf{Keywords: graphene nanoribbons, Raman spectroscopy, length-dependent mode, STM, substrate transfer, vibrational modes, DFT} % remove one length-dep, add "finite size DFT Raman..."

\end{abstract}

\vspace{1cm}
Atomically precise graphene nanoribbons (GNRs) hold the promise of engineering the electronic properties of sp$^2$-carbon systems over a wide range.\cite{Cai.2010, Kimouche.2015, Talirz.2017}
While graphene has several electronic properties that make it an attractive material in certain applications, it lacks an electronic band gap, which limits its use for switching and optoelectronic applications.\cite{Schwierz.2010}
Lateral confinement of charge carriers in narrow graphene nanoribbons is one approach to overcome this limitation.
A major advantage of GNRs is that their electronic band structure can be tailored \textit{via} their width and edge structure. 
GNRs with armchair edges (AGNRs), for instance, cover the full range from quasi-metallic to wide band gap semiconductors.
Based on the number $p$ of carbon-dimer lines present across their width, AGNRs can be grouped into three families: $3p$ (medium gap), $3p+1$ (wide gap) and $3p+2$ (quasi-metallic gap).\cite{Fujita.1996, Son.2006} 
Within each family, the band gap scales inversely with the GNR width. % 
A fabrication approach that allows for atomic control of ribbon width and edge structure is therefore indispensable. 
On-surface synthesis has emerged as the prime technique to fabricate atomically precise GNRs.
Based on metal surface-assisted covalent coupling of specifically designed molecular precursors, this bottom-up approach has proven to be successful for the fabrication of GNRs with different widths (5-,7-,9-,13-AGNRs) and edge topologies (armchair-, zigzag-, chiral-GNRs and GNRs with topological phases).\cite{Cai.2010, Kimouche.2015, Talirz.2017, Chen.2013, Ruffieux.2016, Groning.2018, Rizzo.2018}
%GNRs with zigzag edges (ZGNRs) exhibit spin-polarized localized edge states that are also observed as end-states in finite-length AGNRs.\cite{Fujita.1996, Nakada.1996, Ruffieux.2016, Chen.2013}
In order to exploit GNRs for electronic devices, they need to be transferred from their metallic growth substrate onto a technologically suitable one.\cite{Fairbrother.2017, BorinBarin.2019}
First experiments with GNRs incorporated into devices have shown that the ribbon length as well as their overall integrity are essential for realizing the inherent potential of this material.\cite{Llinas.2017, Chong.2017}
However, methods suited to estimate the length of the GNRs after transfer onto device substrates are lacking. 

% our contribution
Here, we report on a length-dependent vibrational mode in atomically precise AGNRs observed at small Raman shifts. % suitable for the estimation of the GNR length distribution.
This mode, identified as a longitudinal acoustic phonon by first-principle- calculations, is present in all families of AGNRs and is the first vibrational characteristic  that provides information about the GNRs' length after transfer onto device substrates.
We analyze in detail its properties based on a combined study \latin{via} Raman spectroscopy, scanning tunneling microscopy (STM), and computational modelling. 
Finally, we show that the frequency of this mode is highly sensitive to substrate interaction and the presence of defects, making it ideally suited for investigation of GNR integrity and damage monitoring.

\subsection*{Raman Spectroscopy as a tool for GNR characterization}
Due to its ease of use, efficacy, and sensitivity to structural details, Raman spectroscopy has emerged as a primary method for the investigation of GNRs, in particular after transfer onto insulating materials where STM characterization is not possible.
Many of the Raman-active phonon modes in GNRs are named in analogy to the terminology used in graphite, graphene, and carbon nanotubes (CNTs).
For example, the LO-/TO-modes are referred to as the G-peak and its properties have been the focus of several studies.\cite{Gillen.2009, Gillen.2010, BorinBarin.2019} 
In contrast to graphene and CNTs, additional phonon modes found in the spectral range of 1100-1500\tcm~ are not a sign of the presence of defects but rather a result of the (hydrogen-passivated) edges breaking the periodicity of a perfect honeycomb lattice. 
These features, referred collectively as making up the CH/D-region, can be used to identify and probe the edge structure of GNRs.\cite{Vandescuren.2008, Verzhbitskiy.2016} 
The fundamental transverse acoustic mode of GNRs, named radial breathing-\emph{like} mode (RBLM) in reference to its counterpart in CNTs, is commonly used to identify the ribbon width.\cite{Zhou.2007, Vandescuren.2008, Gillen.2010} 
As the Raman shift, intensity, and peak widths associated with these modes depend on the GNR structure, they have been used to monitor the quality of the transfer process and aging of GNRs.\cite{Cai.2010, Fairbrother.2017, Groning.2018, BorinBarin.2019} 
The RBLM, in particular, has been used to monitor growth-related processes such as lateral fusion of GNRs.\cite{Chen.2017, Martini.2019} 

In the above-mentioned studies GNRs are treated as 1D-objects of quasi-infinite length, which allows modelling them in a periodic framework.\cite{Fujita.1996} 
This, however, is often a highly idealized picture as can be seen from STM images showing significant amounts of short ribbons, already present on the growth substrate.\cite{Kimouche.2015}
As the GNRs' length is a crucial factor for their integration into functional devices, a universally applicable method for detecting and modelling finite size GNRs is highly desirable.
In the following, we present such a method which we apply to 5-,\,7-\,and 9-AGNRs, as distinct representatives of all three AGNR families. The GNRs are grown by on-surface synthesis on gold as described by \citeauthor{Cai.2010} and further transferred onto device substrates (see Methods for details).

\subsection*{A longitudinal compressive mode in GNRs}

\begin{figure*}[t!]
\centering
\includegraphics[width=\linewidth]{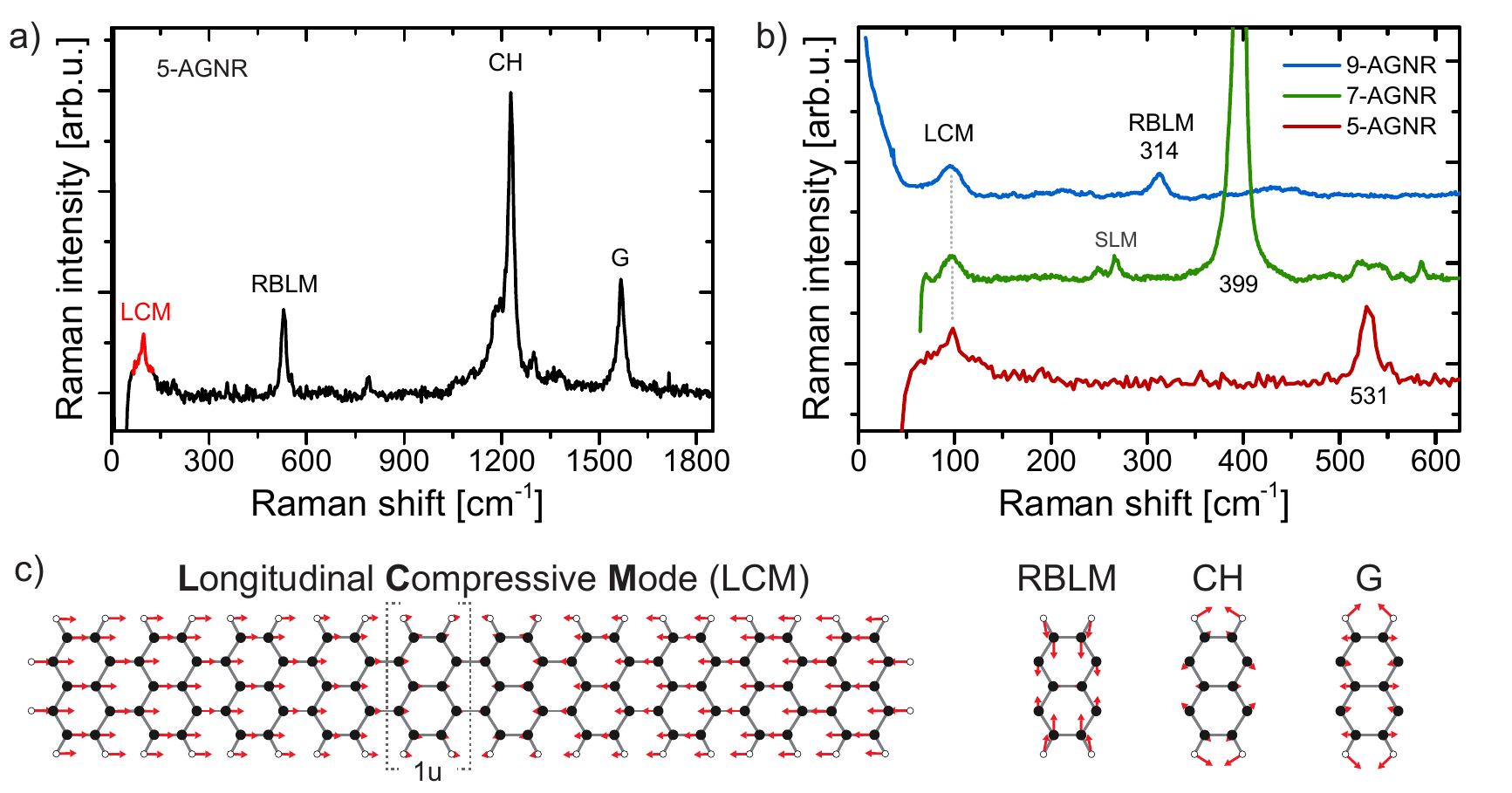}
\caption{\textbf{Longitudinal vibrational mode in armchair GNRs. a)} Raman spectrum of 5-AGNRs with assignment of the most prominent peaks: G-\,\&\,CH- modes, RBLM and a low-energy peak at approximately 100\tcm, labelled LCM. Excitation wavelength \lex = 785\,nm. Sample~5-1\_Au, see Table~\ref{tab:SI_sample_list}. % Sample 5.I.01
\textbf{b)} Low-energy Raman spectra of 5-, 7- and 9-AGNRs (offset for clarity), each exhibiting a peak slightly below 100\tcm. \lex = 785\,nm, 532\,nm and 488\,nm for Samples~5-1\_Au, 7-1\_T, 9-1\_T, respectively. % Samples 5.I.01, 7.Br.01, 9.I.06
\textbf{c)} Atomic displacement profiles obtained from DFT calculations of 5-AGNRs matching the peaks in a). LCM for a finite-size 10-unit GNR. A single naphthalene unit is referred to as 1u. RBLM, CH- and G-modes are shown for a single unit-cell of a periodic ribbon. C-atoms are shown with filled, H-atoms with open circles.}
\label{fig:Fig1}
\end{figure*}

Figure~\sfigref[a]{fig:Fig1} shows a Raman spectrum of 5-AGNRs obtained directly on the Au(788) growth substrate. 
The corresponding STM images are shown in Figure~\ref{fig:FigSI_STM_bare_Au788_5.I.01}. 
Unless stated otherwise, Raman spectra are obtained in vacuum using an optimized mapping approach to minimize damage to the GNRs, to obtain a high signal-to-noise ratio, and to probe the average properties of the GNR film.\cite{Overbeck.2019}
In the spectrum we identify the G-, CH- and RBLM-peaks described above. 
In addition to these modes, we detect a low-energy peak slightly below 100\tcm (labelled LCM).
This mode has, to the best of our knowledge, not been discussed for GNRs to-date and is the focus of this study. We argue below that it originates from a longitudinal vibration of the GNR and that it is extremely useful for GNR characterization and to probe their structural integrity upon transfer and device integration.

We first discuss the low-energy modes that have already been reported for armchair-type GNRs and how they can be distinguished from this new mode. The lowest mode usually reported is the well-documented RBLM, the fundamental acoustic mode for which all atoms move in-plane in the direction perpendicular to the ribbon axis (see Figure~\sfigref[c]{fig:Fig1}).
It is observed at a Raman shift of about 534\tcm, 399\tcm~ and 313\tcm~ for the 5-,\,7-\,and 9-AGNRs, respectively. Modes at energies below the RBLM of a particular GNR have been attributed to the formation of wider GNRs with correspondingly lower RBLM frequency by thermally-induced lateral fusion.\cite{Chen.2017, Martini.2019}~
We can exclude this effect here as we do not reach the temperatures required to induce lateral fusion, nor do we observe them in STM or measure the characteristic series of RBLMs for GNRs of integer multiple widths of the fundamental ribbon.\cite{Chen.2017}
Another low-energy mode is the shear-like mode (SLM), which has been observed in 7-AGNRs and has recently been discussed for 5-AGNRs and 9-AGNRs. \cite{Ma.2017, Overbeck.2019}

To clarify whether the new LCM-mode at 100\tcm~ is specific to 5-AGNRs or is a universal property of all AGNRs, we also acquired Raman spectra on 7-\,\&\,9-AGNRs.
Figure~\sfigref[b]{fig:Fig1} displays the low-energy spectra of 5-, 7-, and 9-AGNRs. In all spectra, a peak at roughly the same Raman shift is clearly identified, pointing towards the universal presence of this mode in AGNRs. 
While the spectrum for 5-AGNRs was acquired directly on the Au(788) growth substrate, the spectra for 7-\,\&\,9-AGNRs were acquired after transferring the GNRs to a device substrate optimized to enhance the Raman signal.\cite{Overbeck.2019} 
For each type of GNR the wavelength and laser power were independently optimized to obtain the highest signal to noise ratio.
The presence of this mode in multiple AGNRs, always at frequencies below their fundamental transverse acoustic mode (RBLM) points towards an even longer-range mode associated with a longitudinal acoustic vibration.

First-principles modelling of finite-size GNRs, in contrast to the usual approach of simulating infinitely long ribbons \latin{via} periodically repeated unit cells, produces precisely such a longitudinal compressive mode (LCM) (see Figure~\ref{fig:Fig1}c).
To better understand the nature of this low-energy mode, we performed Raman spectra calculations using both density functional theory (DFT) as well as separate calculations using a combination of the REBOII force-field and a bond polarizability model. The latter calculations were carried out in order to examine overall trends and larger systems not easily tractable by DFT. For details see Methods and SI Note~2.

\subsection*{Length-dependence and quantum mechanical calculations}

\begin{figure*}[t!]
\centering
\includegraphics[width=\textwidth]{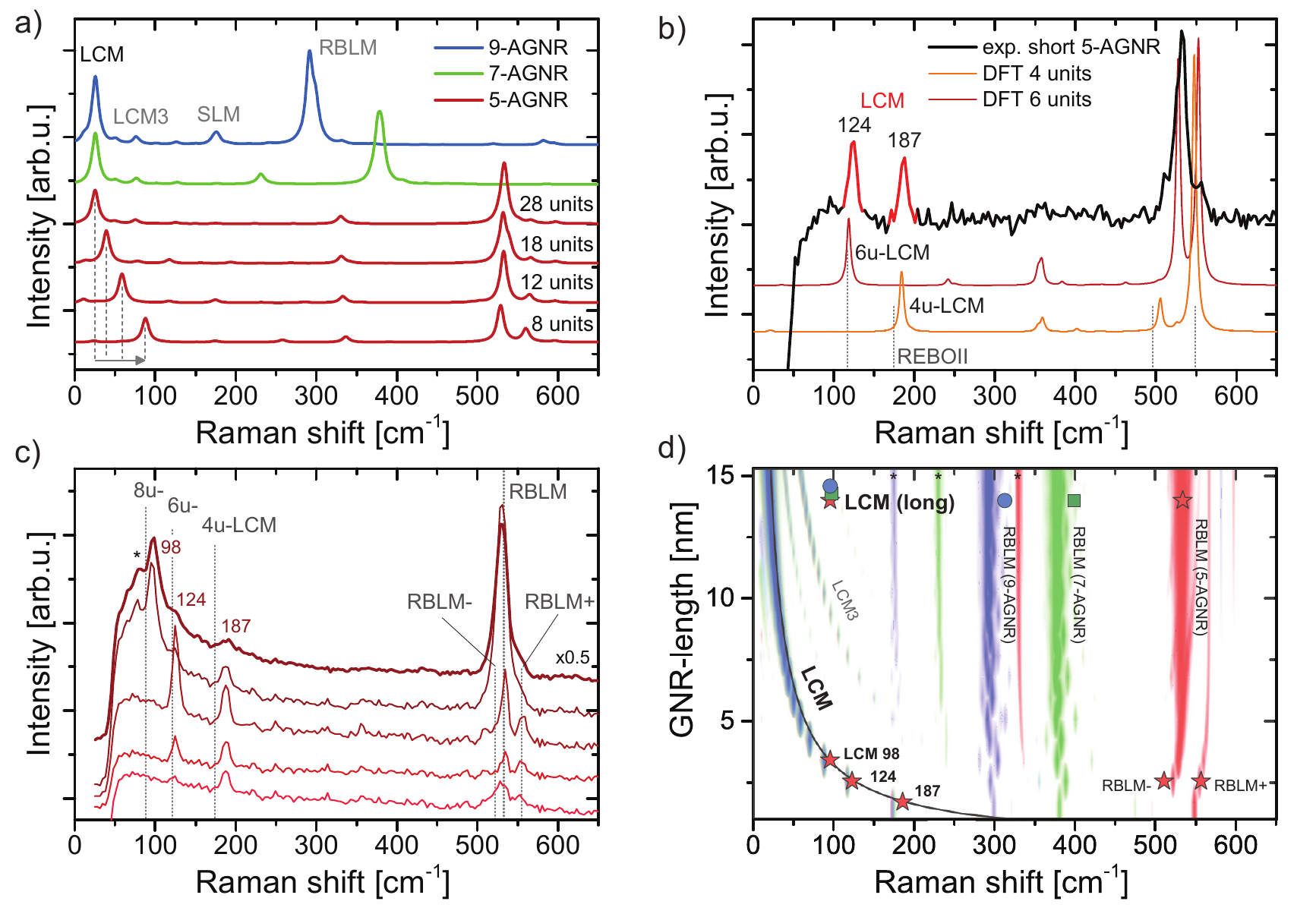}
\caption{\textbf{Length-dependence of GNR-modes in experiment and Raman calculations. a)} REBOII-based spectra for long, finite (28\,units $\approx$ 12\,nm length) 9-, 7-, and 5-AGNRs and shorter 5-AGNRs. 
\textbf{b)} Experimental Raman spectrum for 5-AGNRs from a sample with high content of short ribbons (black line) and DFT-calculated spectra for 4- \& 6-unit (4u, 6u) length 5-AGNRs. Dotted lines indicate the frequencies obtained from the REBOII-based calculations for the corresponding LCM modes. \lex = 785\,nm, Sample~5-6\_Au. %Sample 5.Br.02 
\textbf{c)} Raman spectra for a selection of 5-AGNR samples on Au substrates, offset for clarity. 
REBOII-calculated frequencies for the 4/6/8u-LCM and 6u-RBLM\textsuperscript{+/-} are indicated in grey. \lex = 785\,nm, Samples~5-1\_Au to 5-5\_Au, top to bottom. % Samples 5.I.04, .07, .08, .06, .01 bottom to top
\textbf{d)} Map of REBOII-calculated length-dependent spectral intensity, superimposed with the experimental mode frequencies (symbols) and analytical model (black line). Intensity values are color-coded on a logarithmic scale from white to red (green/blue) for 5-AGNRs (7-/9-AGNRs). Symbols are colour coded accordingly. The LCM for the three GNRs overlap, preventing distinction by color. Black asterisks indicate the shear-like modes (SLM).
%The length-dependence of the LCM (and overtones) is clearly visible, while RBLM and SLM (indicated with grey arrows) are roughly independent of length from about 10nm.
}
\label{fig:FigTheo3}
\end{figure*}

In Figure~\sfigref[a]{fig:FigTheo3}, we initially compare the calculated Raman spectra for all three GNR widths for a length of about 12\,nm (i.e. 28 units). For these long GNRs, we employ the combined REBOII and bond polarizability method as Raman calculations at the DFT level are prohibitively computationally expensive.

The most prominent peak is the RBLM, which shows the well-known width-dependent Raman shift.\cite{Zhou.2007, Vandescuren.2008, Gillen.2010}
Also, the position of the shear-like mode (SLM) mentioned earlier shows a clear down-shift with increasing GNR width.\cite{Ma.2017}%, Liu.2019}
In contrast, the frequency of the LCM and its overtones does not change as a function of width of the three GNRs, matching our experimental observations in Figure~\sfigref[b]{fig:Fig1}.
This can be rationalized by the nearly constant ratio of mass per unit cell to the number of bonds along the GNR axis. 
In a simple mass-and-spring approximation, this ratio determines the frequency of the normal mode considering the atomic displacement profile of the LCM along the GNR axis, as shown in Figure~\ref{fig:Fig1}c and \ref{fig:FigSI_ModeProfiles}.
It would therefore be interesting to investigate GNRs with armchair segments of different width to clarify the effect of changing the vibrating mass vs. ribbon stiffness.\cite{Rizzo.2018, Groning.2018}~ 
%, such as those reported recently by \citeauthor{Rizzo.2018, Groning.2018} in order to gain deeper understanding in the relationship of vibrating mass vs. ribbon stiffness when tuning the number of 7-\,vs.\,9-atom wide segments.
%\todo{Theoretical analysis 7-9-7 or 7-11-7AGNR, maybe experiment? Gabriela has ribbons ready!}

The calculations show a strong change of the LCM frequency as a function of GNR length. In the lower part of Figure~\sfigref[a]{fig:FigTheo3} we show three spectra of shorter 5-AGNRs that exhibit a progressive up-shift of the LCM as the ribbon gets shorter.
To probe this effect experimentally, we fabricated a sample with an unusually high number of very short 5-AGNRs as determined from low temperature STM (LT-STM; see Figure\,\ref{fig:FigSI_STM_LT-short5}).
Short GNRs are usually a result of a premature hydrogen-passivation during the polymerization step, or can be caused by mono-bromo(iodine) functionalized molecules,\cite{DiGiovannantonio.2018} which can be used to identify inappropriate growth conditions or inadequate storage of the precursor molecules.
% Prolonged storage of the precursor molecules or inappropriate growth conditions can lead to a large fraction of very short ribbons, because polymerization of the precursor molecules is stopped as soon as a radical is missing, making this relevant from both a fundamental as well as technological perspective.\cite{Kimouche.2015} 
%\todo{Decide if and where to include Martini.2019}
The shortest ribbon segments obtained in our experiments with 5-AGNRs are 4-units and 6-units long, composed of two and three precursor molecules, respectively.
Such short ribbons are extremely mobile at room temperature requiring low-temperature STM characterization in order to quantify the GNRs' length distribution (see Figure~\ref{fig:FigSI_STM_LT-short5}).
%\todo{Add RT-STM image with fuzzy area?} 
Figure~\sfigref[b]{fig:FigTheo3} displays the Raman spectrum of a sample with short 5-AGNRs, exhibiting two prominent peaks below 200\tcm.
Also plotted are the calculated Raman spectra for 4- and 6-unit 5-AGNRs, which can be treated entirely in DFT because of their limited size. 
The spectra match well with the peak positions observed experimentally for the LCM, without any need for frequency scaling.
Moreover, the DFT-spectra also reproduce the side-peaks observed on the RBLM, which result from the splitting of the RBLM into normal modes with diagonal atomic displacement for these short ribbons (see Figure~\ref{fig:FigSI_ModeProfiles} for the atomic displacements).
To assess the sensitivity of the Raman spectra on the computational method, we also calculated the LCM frequencies \latin{via} forces calculated by the REBOII potential (dotted lines). 
The results are very similar, validating its use for longer GNRs. The results also indicate that the relevant REBOII-calculated forces tend to be weaker than those from DFT and thus the mode frequencies are slightly underestimated for these acoustic modes (also observed for the RBLM of 7-/9-AGNRs).

Next, we experimentally investigate the influence of the ribbon length.
Figure~\sfigref[c]{fig:FigTheo3} shows Raman spectra obtained on several samples of 5-AGNRs with changing percentages of short ribbons tuned \latin{via} the growth parameters. The spectra are sorted, from bottom to top, by increasing average GNR length as deduced from STM imaging.
The bottom spectrum shows the peak at 187\tcm ~attributed to the 4u-LCM, as well as a small peak at 125\tcm ~where the 6u-LCM is expected. 
For increasing average ribbon length, the relative amplitude of the 6u-LCM becomes more prominent, whereas the 4u-LCM decreases. 
For even longer average ribbon length, a third peak appears centred at 98\tcm, which we attribute to the 8u-LCM, while the peaks at 125\tcm~and 187\tcm~ disappear.
Notably, unassigned modes at comparable frequencies can be found in Raman spectra published by other groups.\cite{Martini.2019, MedinaRivero.2019} 
%\todo{careful phrasing.}
The RBLM shows a similar trend with side peaks clearly visible for the same spectra that show the strongest signal of 4u-\,\&\,6u-LCM. For the two topmost spectra, the central RBLM-peak characteristic of long GNRs is recovered.

Figure~\sfigref[d]{fig:FigTheo3} provides an overview of the theoretical and experimental data in a single plot. 
For the calculations, the length of the GNRs is varied from 4 - 36 units (i.e., 1.7\,nm to 15.3\,nm) along the y-axis. 
The computed Raman scattering intensities are represented on a logarithmic color-scale from white to red for 5-AGNRs, and from white to green/blue for 7-/9-AGNRs as a function of Raman shift (x-axis).
As shown in Figure~\sfigref[a]{fig:FigTheo3}, the RBLMs of the three GNRs are well separated and do not significantly change as a function of length, except for the splitting seen in very short 5-AGNRs.
The LCM, in contrast, overlaps for the three different GNRs and shows a common frequency-downshift with increasing GNR length.
This trend is also seen for its overtone, labelled LCM3 on the figure.
The SLMs are marked by small asterisks and are essentially independent of length, as expected.
The experimental data are superimposed as symbols with the same red/green/blue color-coding for 5-/7-/9-AGNRs, respectively.
We display the experimental data at an assigned length of 14-15\,nm for `long' GNRs produced with the optimized growth parameters. This is a lower bound for the length distribution determined from STM images (see Figure~\sfigref[b]{fig:FigSI_Halogens}).\cite{DiGiovannantonio.2018} 
The discrepancy in Raman shift of the LCM between experiment and theory for these long ribbons is addressed in the Discussion section below.
Finally, for the LCM we plot an analytical curve using a model derived by \citeauthor{Gillen.2010}~to predict the frequency of the RBLM as a function of width.
This frequency $\omega_{\text{LCM}} = a\cdot \pi/l$, where $l$ is the ribbon length and $a = \SI{1026}{\angstrom\per\centi\meter}$ is the slope of the LA-phonon in graphene, is in excellent agreement with our other calculations as well as our experiments.\cite{Gillen.2010b}%, Liu.2019} 

\begin{figure*}[t!]
\centering
\includegraphics[width=\linewidth]{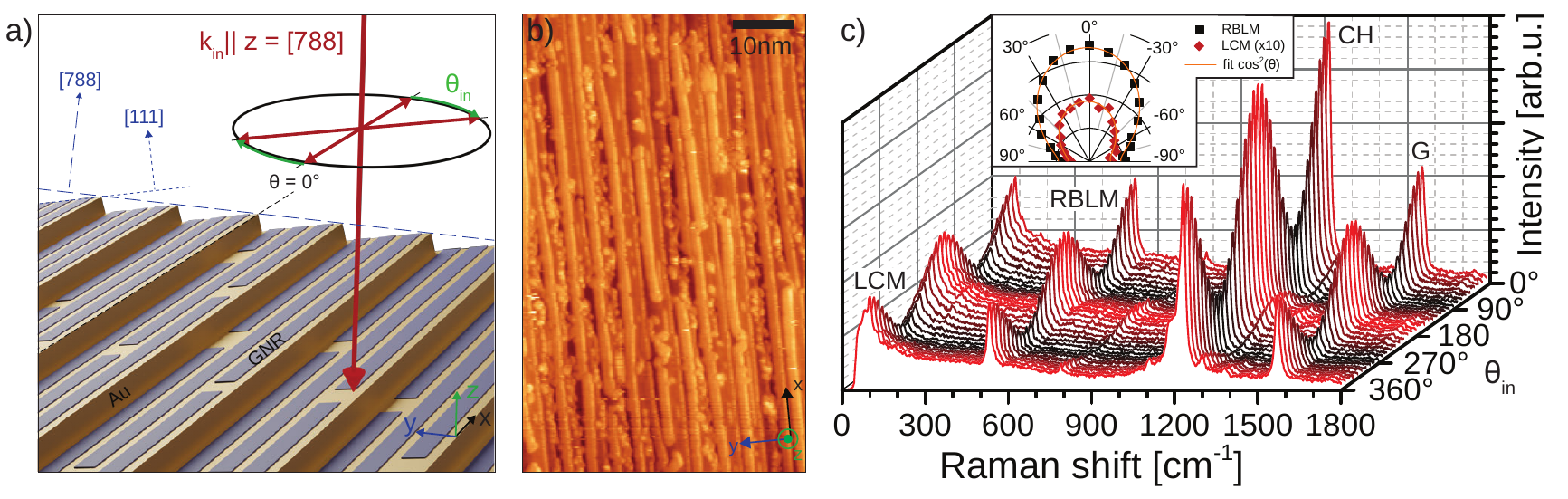}
\caption{\textbf{Characterization of the LCM for 5-AGNRs. a)} Sketch of the measurement geometry for aligned GNRs grown on terraced Au(788) crystals. The incoming light under normal incidence to the Au(788) surface is polarized at an angle $\theta_{\text{in}}$ relative to the direction of the gold terraces. \textbf{b)} STM image of aligned GNRs (sparsely grown 9-AGNRs for the purpose of illustration). \textbf{c)} Raman spectra of aligned 5-AGNRs on a Au(788) growth substrate as a function of polarization of the exciting laser at \lex = 785\,nm. Sample~5-1\_Au. Inset: polar plot of the RBLM \& LCM (scaled x10) mode intensities $I \propto \text{cos\textsuperscript{2}}(\theta)$.%, with a polarization anisotropy of $\text{P\tsub{LCM,(RBLM)}} = 0.5, (0.6)$, see~SI Note~3. %\ref{SI: PolarizationExpl.}. % Sample 5.I.01
}
\label{fig:FigPol2}
\end{figure*}

\subsection*{Polarization analysis of the LCM}

Polarization-dependent measurements provide additional information on the nature of Raman-active vibrations.
As a consequence of their high aspect ratio, GNRs show a characteristic anisotropic dependence of scattering intensity on the relative alignment of light polarization direction and ribbon axis, an effect known from other quasi-1D materials such as carbon nanotubes.\cite{Gommans.2000,Wang.2001,Jorio.2002, Saito.2010, Senkovskiy.2017} 
We probe this dependence for 5-AGNRs grown on a Au(788) crystal where the narrow (3-4\,nm) (111)-terraces favor unidirectional growth of 5-AGNRs parallel to the step edges separating the terraces (see~Figure~\sfigref[a, b]{fig:FigPol2}).
To probe the global alignment of GNRs by Raman spectroscopy we vary the angle $\theta_{\text{in}}$ of the linear polarization of the excitation laser with respect to the direction of the terraces $\theta=\SI{0}{\degree}$ (see Figure~\sfigref[a]{fig:FigPol2} and SI-Note~3). 
No polarizer was used in the detection path to maximize the signal intensity.
Figure~\sfigref[c]{fig:FigPol2} shows a waterfall plot of Raman spectra of aligned 5-AGNRs as a function of polarization angle.
The intensity of the low-energy LCM clearly follows that of the RBLM, CH- \& G-modes.
The inset shows the peak intensities of the RBLM and LCM displayed in a polar plot. The behaviour is well-described by an ensemble of dipoles, each with an $\text{I}(\theta_{\text{in}}) \propto \text{cos}^2(\theta_{\text{in}})$ characteristic, that are preferably oriented along the $\theta=\SI{0}{\degree}$-direction.\cite{Senkovskiy.2017} 
The fact that all GNR modes, including the LCM, follow a common polarization dependence irrespective of intrinsic mode symmetry is attributed to the strong shape anisotropy.\cite{Jorio.2002, Saito.2010} 
We use the polarization anisotropy P as a measure to quantify this effect and find a value of P\tsub{LCM,(RBLM)} = 0.5, (0.6) see~SI Note~3. %\ref{SI: PolarizationExpl.}
The length of GNRs also affects the observed polarization-dependence.
In Figure~\ref{fig:FigSI_PolDep_short5} we show measurements on samples containing predominantly short GNRs. 
We observe a lower polarization anisotropy P\tsub{RBLM} = 0.4 than for the GNRs grown with optimized parameters (Figure~\sfigref[c]{fig:FigPol2}). 
We attribute this to the combination of two effects: First, a lower degree of ribbon alignment as seen in STM because short ribbons are not restricted in growth direction by the width of the terraces, and second, a reduced shape-anisotropy as reflected in the optical absorption matrix elements.\cite{Saito.2010}~

Finally, we investigate the effect of transferring the GNRs from their gold growth substrate to standard silicon and Raman optimized device substrates.\cite{Overbeck.2019} 
Figures~\ref{fig:FigSI_BG_sub}, \ref{fig:FigSI_5-before_after}, and \ref{fig:FigSI_extra9} of SI Note~3~show the LCM in spectra for the three types of AGNRs after transfer. Again, we performed polarization-dependent measurement on the LCM of aligned 7- and 9-AGNRs. 
As for aligned 5-AGNRs, the LCM follows the polarization dependent intensity of the other Raman active GNR modes (Figure~\ref{fig:FigSI_PolDep_7-9}).
For 9-AGNRs with an average length of about 40\,nm we observe a significantly larger polarization anisotropy ($\text{P\tsub{LCM}} > 0.9$ and $\text{P\tsub{RBLM}} > 0.7$).
Samples of (globally) non-aligned 5-AGNRs synthesized on Au/Mica substrates with wider terraces, too, show the LCM and RBLM at the same frequencies but without any polarization dependence (see SI Note~1 %\ref{SI: Note 1 - fab and STM} 
and Figures~\sfigref[b]{fig:FigSI_Halogens}~and \sfigref[a]{fig:FigSI_5-before_after} for STM and Raman spectra). 
Beyond confirming the ribbon-origin of the LCM, these polarization-dependent measurements can therefore be used as an independent, albeit rough, indication of GNR length for well-aligned ribbons.
Altogether, for all three families of AGNRs, we find a common polarization dependence of the LCM and the other Raman modes, and that the transfer procedure does not significantly impact the GNR structure as revealed by its vibrational modes.

\begin{figure*}[t]
    \centering
    \includegraphics[width=\textwidth]{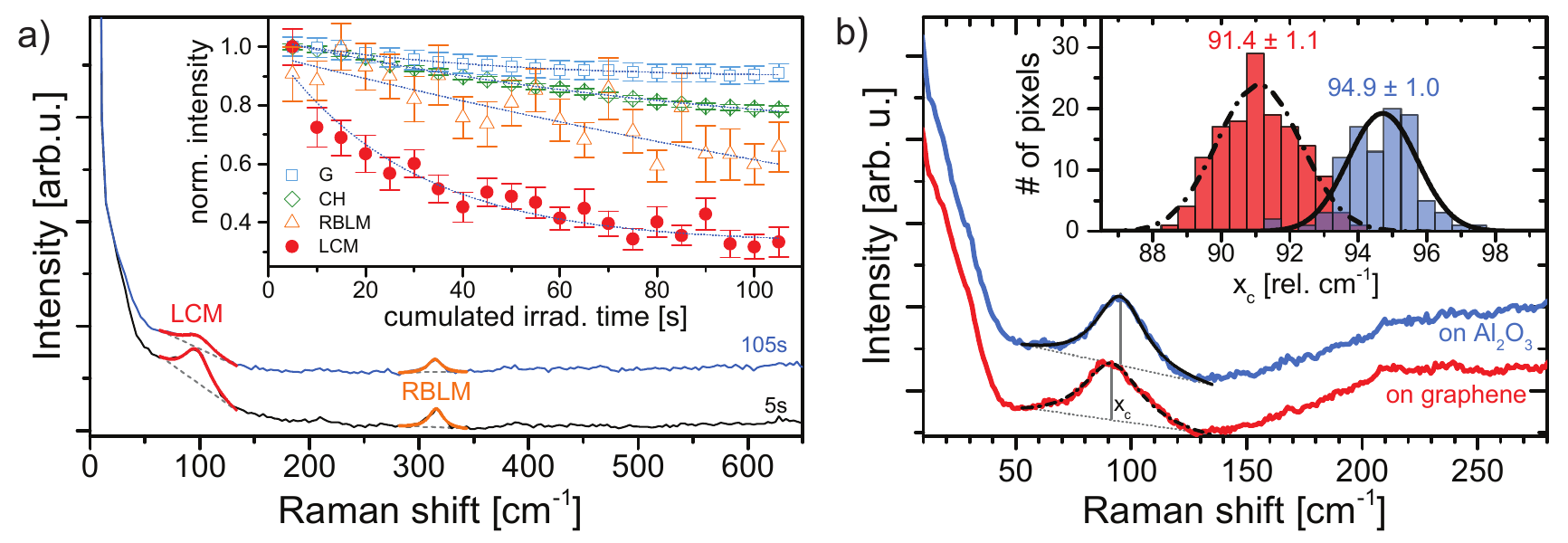}
    \caption{\textbf{Damage monitoring and substrate sensitivity \latin{via} the LCM-peak of 9-AGNRs. a)} Low energy spectra acquired with 5\,s integration time and after an additional 100\,s laser irradiation time at the same sample location, showing GNR damaging. The inset shows the normalized peak intensity obtained from a Lorentzian fit of the LCM, RBLM, CH- \& G-peaks as a function of the cumulated irradiation time. Full data in Figure~\ref{fig:FigSI_Damage_full}. Dashed lines are fits with simple exponential decays. \lex = 488\,nm, $P=\SI{5}{\milli\watt}$, Sample~9-2\_T. % Sample 9.I.07
    \textbf{b)} Substrate-dependent Raman shift of the LCM for GNRs on oxide or on a graphene inter-layer. The inset shows histograms for the peak center extracted from a single Raman map containing both substrates, fitted with Gaussian distributions. \lex = 488\,nm, Sample~9-1\_T. % Sample 9.I.06
    }
    \label{fig:Fig4}
\end{figure*}

\subsection*{LCM as a tool to assess ribbon damage and substrate interaction}

In addition to being useful to rapidly assess non-ideal growth regimes that result in short GNRs, the LCM is also a sensitive probe for ribbon integrity. 
The atomic displacement shown in Figure~\sfigref[c]{fig:Fig1} shows that the LCM requires the coherent vibration of all atoms along the ribbon. 
It can therefore be expected that the presence of defects in GNRs as well as the interaction with the substrate on which the ribbon is placed has a strong influence on it, especially for long ribbons.

To systematically investigate damaging of GNRs, a transferred film of 9-AGNRs was irradiated with high laser power (5\,mW) at 488\,nm, leading to the introduction of defects.\cite{Senkovskiy.2017} 
Figure~\sfigref[a]{fig:Fig4} shows spectra acquired on the same sample spot after short and prolonged laser exposure. 
The inset shows the fitted intensity of the LCM, RBLM, CH- and G-peaks as a function of the accumulated irradiation time by the excitation laser while acquiring the spectra. While all peaks show an intensity decrease with increasing irradiation time, this decay is most pronounced for the LCM. 
Figure~\sfigref[c, d]{fig:FigSI_Damage_full} displays a power-dependent series of spectra, all normalized to a common power-time product. The LCM clearly disappears as the power and corresponding laser damage is increased.
This shows that the LCM intensity is a good tool for monitoring the ribbon integrity throughout device processing and may be a key identifier for GNRs' suitability in optoelectronic applications.

Inspired by the long-range nature of this mode, we investigated whether there is any influence of the target substrate onto which the GNRs are transferred.
Two technologically relevant substrates are oxide surfaces used for gate-insulation in field-effect transistors and graphene, which is increasingly used as an electrode material.\cite{ElAbbassi.2019b, Martini.2019}
We transferred a sample of 9-AGNRs on a patterned graphene-on-oxide substrate and acquired Raman spectra scanning over the boundary of both substrates (Figure~\ref{fig:FigSI_Substrate-Interaction}). Interestingly, we find a strong down-shift of the LCM frequency for GNRs on graphene, a change of nearly 4\%. No comparable shift is observed for any of the other Raman modes of GNRs.
Low-frequency modes have been used extensively over the last years to probe the layer interactions of 2D-materials.\cite{Tan.2012, Huang.2016, Liang.2017}
The observation of a significant substrate-dependent effect on the LCM points towards a host of so far unexplored phenomena relevant for the fabrication of hybrid devices that can be explored \textit{via} these low-frequency modes.

\subsection*{Discussion}

We find a good overall match between Raman spectra, first-principles calculations and independent observations from (low-temperature)-STM.
The high background and low Raman intensity when measuring on gold, however, prevent us from observing the LCM in the samples containing on average the longest GNRs while still on the growth substrate (see Figure~\ref{fig:FigSI_extra9}). 
Moreover, the frequency of the LCM observed after transfer to device substrates is substantially higher than that predicted by calculations for isolated GNRs, matching better with the overtone denoted LCM3.
In contrast, for short 5-AGNRs the LCM is observed precisely at the frequencies predicted by DFT on both growth and device substrates.
To address this discrepancy for long GNRs we focused on 9-AGNRs, which reproducibly show the LCM at 96\tcm~ for \lex=488\,nm, the wavelength for which our setup allows measurements at the lowest wavenumbers. 
The spectra in Figure~\sfigref[b]{fig:FigSI_PolDep_7-9} show a clear polarization dependence not only of the LCM at 96\tcm~ but also of the background below about 60\tcm.
Our spectral resolution still does not allow us to discriminate whether this low-energy polarization dependent background is a signature of another lower-energy Raman peak or due to increased scattering of the tail of the laser from the ribbons, which should also follow the variation of the GNRs' interaction cross-section with polarization. 
Selective synthesis of intermediate length (5-7\,nm) GNRs would allow to find a definite answer.
Based on the strong substrate-sensitivity shown in Figure~\sfigref[b]{fig:Fig4}, we interpret the observed peak at 96\tcm~ as the LCM of 9-AGNRs being up-shifted in energy relative to the gas phase calculations due to the interaction with the substrate. 
In particular, we attribute this shift to localized pinning of the GNR on the rougher device substrate. 
This effect, which is not included in our calculations, can lead to a reduced effective length of freely vibrating GNR sections thereby up-shifting the modes towards the experimentally observed frequencies.
The importance of modelling the substrate is further highlighted by its selective influence on longer GNRs.
To accurately model an experimental spectrum, which includes contributions of typically 100s of GNRs within the size of the laser spot, one can extract the length distribution of GNRs from high-resolution STM images and compute an accordingly weighted sum of calculated spectra (see Figure~\ref{fig:FigSI_SummedMD} in SI Note~4).
This summation approach qualitatively reproduces the spectra, in particular the signatures of short ribbons and the observed increase in background intensity around 100\tcm. 
The discrepancy with the experimental spectra is, again, attributed to the fact that we neglect the substrate effect in the calculations.
We anticipate that the inclusion of such a substrate model will allow the reverse approach of deducing a quantitative length-distribution of GNRs in a sample by Raman spectroscopy alone.
An improved understanding of substrate effects is the focus of an ongoing study and beyond the scope of this paper.

\subsection*{Conclusion}

We have identified and characterized a so-far unreported longitudinal vibrational mode in AGNRs by Raman spectroscopy.
Their length-dependent Raman shift has been investigated by adjusting the GNR growth parameters and is supported by a combined analysis with high resolution STM and first-principles calculations. 
% Employing molecular dynamics and density functional theory, we modelled its length-dependence and identified the atomic displacements associated with the excitation of this mode.
We demonstrated the usefulness of the LCM for the rapid identification of samples containing short ribbon-segments on both growth and device substrates. This information would otherwise require time-consuming and expensive low-temperature STM for probing the GNRs' length on the growth substrate and would not at all be accessible after GNR transfer and device integration. Finally, we show the mode's sensitivity to damage of the GNRs that makes it ideally suited to study the effects of processing and nano-fabrication on GNRs, which therefore has become a routine practice in our labs.
Moreover, we observe a remarkable influence of the substrate on the vibrational frequency of this mode, pointing towards the possibility to investigate the interaction of GNRs with their environment by Raman spectroscopy.
% We expect that due to its long wavelength this mode will show a strong dependence on interaction with the underlying substrate. First hints towards this effect are the observation that this mode is extremely hard to observe for the case of long ribbons in close contact with the catalytic metal growth substrate.
%\todo{Plot 9-AGNR high res before and after transfer to compare S/N \& LCM visibility.}
% While the RBLM is useful to verify the GNR-width known \emph{a-priori} through the choice of molecular precursors, we believe that the LCM can be used to identify a host of hitherto unexplored aspects of GNR.
We anticipate that this mode will also be observed in GNRs with different edge structures and that these findings will substantially advance the overall understanding of GNRs, their interaction with diverse substrates, and provide insights that we deem crucial for the deterministic fabrication of GNR-based devices.

\section*{Methods}

\textbf{On surface synthesis and STM characterization of AGNRs.} 
5-AGNRs were synthesized from an isomeric mixture of 3,9-diiodoperylene and 3,10-diiodoperylene (DIP) as the precursor monomer - referred to as iodine-based. 
5-AGNRs with bromine-based precursors were synthesized from an isomeric mixture of  3,9-dibromoperylene and 3,10-dibromoperylene (DBP).\cite{Schlichting.1997}  
7-AGNRs were synthesized from 10,10'-dibromo-9,9'-bianthryl (DBBA)\cite{Cai.2010} and 9-AGNRs from 3',6'-di-iodine-1,1':2',1''-terphenyl (DITP).\cite{DiGiovannantonio.2018} 
First, Au(788) single crystals (\textit{Matec}, Germany) or Au/mica substrates (\textit{Phasis}, Switzerland) were cleaned in ultra-high vacuum by two sputtering/annealing cycles: 1 kV Ar+ for 10 minutes followed by annealing at $\SI{470}{\celsius}$ for 10 minutes. Next, the monomer was sublimed onto the Au surface from a quartz crucible heated to $\SI{70}{\celsius}$ (DITP), $\SI{200}{\celsius}$ (DBBA), and $\SI{200}{\celsius}$ (DBP/DIP) respectively, with the substrate held at room temperature. 
After 1 monolayer coverage deposition, for both 7-, and 9-AGNRs the substrate was heated (0.5\,K/s) up to $\SI{200}{\celsius}$ with a 10 minute holding time to activate the polymerization reaction, followed by annealing at $\SI{400}{\celsius}$ (0.5\,K/s with a 10 minute holding time) in order to form the GNRs \latin{via} cyclodehydrogenation. For the synthesis of the 5-AGNRs a slow annealing (0.2\,K/s) was carried up to $\SI{225}{\celsius}$.

%\textbf{Scanning tunneling microscopy (STM):}
Topographic scanning tunneling microscopy images of as-grown AGNRs on Au(788) and Au/mica were taken with a \textit{Scienta Omicron} VT-STM/LT-STM operated at room temperature/5\,K. Constant-current STM images were recorded with -1.5\,V sample bias and 0.03\,nA setpoint current. %low-temperature STM exclusively in supplementary.

\textbf{Transfer of GNRs to substrates.}
AGNRs were transferred from their Au/mica or Au(788) growth substrate to silicon-based substrates by two different transfer approaches. Au/mica samples were transferred using a polymer-free method described elsewhere.\cite{BorinBarin.2019} 
Samples grown on Au(788) crystals were transferred by an electrochemical delamination method. First, PMMA was spin-coated (2500\,rpm for 90 seconds, 4 layers) on GNR/Au to act as a support layer during the transfer, followed by a 10 minutes curing step at $\SI{80}{\celsius}$. 
In order to decrease the PMMA delamination time, the PMMA on the edges of the Au (788) crystal was removed after UV-exposure (80 minutes) followed by 3 minutes development in water/isopropanol. The electrochemical delamination transfer was carried out using 1\,M NaOH aqueous solution as electrolyte. The electrochemical cell was mounted using a carbon rod as anode and the PMMA/GNR/Au as the cathode. A DC voltage of 5V was applied between anode and cathode for 1 minute (resulting current around 0.2\,A). 
Hydrogen bubbles form at the interface of PMMA/GNR and Au, resulting in delamination of the PMMA/GNR layer. The PMMA/GNR layer was cleaned for 5 minutes in purified water before being transferred to the target substrate. Subsequently, the PMMA/GNR/substrate stack was cured for 10 minutes at $\SI{80}{\celsius}$~followed by 20 minutes at $\SI{110}{\celsius}$~to increase the adhesion between the target substrate and the PMMA/GNR layer. Finally, the PMMA was dissolved in acetone (15 minutes) followed by an ethanol-rinse.

\textbf{Raman spectroscopy experiments.}
Raman spectra were acquired with a \emph{WITec Alpha 300 R} confocal Raman microscope in backscattering geometry. 
The linear polarization of the exciting lasers was adjusted with a motorized $\lambda/2$ plate. The backscattered light was filtered with an analyzing polarizer only where explicitly stated and coupled to one of two spectrometers: a 300\,mm lens-based spectrometer with gratings of 600\,g/mm or 1800\,g/mm equipped with a thermoelectrically cooled CCD for 488\,nm and 532\,nm excitation and a 400\,mm lens-based spectrometer with gratings of 300\,g/mm or 1200\,g/mm equipped with a cooled deep-depletion CCD for 785\,nm excitation.
The laser wavelength, power and integration time were optimized for each type of GNR and substrate to maximize the signal.\cite{Overbeck.2019}
The sample was mounted in a home-built vacuum chamber at a pressure below 10\textsuperscript{-2}\,mbar, mounted on a piezo stage for scanning.
A polynomial background was subtracted from the raw spectra to remove signatures of photoluminescence (see Figure~\ref{fig:FigSI_BG_sub}).\cite{Senkovskiy.2017}

\textbf{Computational methods.}
Normal modes and Raman intensities were calculated using density functional theory (DFT) for small finite size and periodic GNRs, and  a combination of a force-field and bond polarizability model for larger finite-size systems. For the smaller systems, DFT calculations were performed with the VASP program\cite{Kresse.1993, Kresse.1996, Kresse.1996b} with projector-augmented-wave pseudopotentials\cite{Kresse.1999} and the Perdew–Burke–Ernzerhof exchange-correlation functional.\cite{Perdew.1996} A plane-wave cutoff of 600\,eV was used and prior to other calculations all structures were relaxed in VASP until residual forces were less than $\SI{e-4}{e\volt\per\angstrom}$ in magnitude. Resonant Raman intensities were calculated by the Placzek approximation\cite{Long.2002} where the derivatives of the DFT-calculated frequency dependent dielectric matrix\cite{Gajdos.2006} obtained from a finite difference method.\cite{Liang.2014} Larger systems were treated with a combination of the REBOII potential\cite{Brenner.2002} for forces and a bond polarizability model\cite{Saito.2010} for Raman intensities. The bond polarizability parameters used were $\alpha_\parallel - \alpha_\perp=\SI{0.32}{\angstrom^3}$, $\alpha'_\parallel + 2\alpha'_\perp = \SI{7.55}{\angstrom^2}$, and $\alpha'_\parallel - \alpha'_\perp = \SI{2.60}{\angstrom^2}$.\cite{Nakadaira.1997,Saito.2010} In both cases, the dynamical matrix was calculated using the finite difference method with a displacement of \SI{0.03}{\angstrom} \latin{via} the use of the phonopy program package\cite{Togo.2015} in combination with forces calculated by either DFT or REBOII. Intensities are calculated as an average over all backscattering geometries, spectra are then plotted as a sum of Lorentzians with a width of 10\tcm.

\section*{Author contributions statement}
J.O. performed Raman measurements; J.O., G.B.B.; M.P. M.D.L. and M.C. analyzed the Raman results;  C.D. performed Raman calculations; G.B.B. optimized the GNR growth; G.B.B, R.D. and Q.S performed on-surface synthesis and STM imaging of GNR; J.O., O.B. fabricated device substrates and with G.B.B. and R.D. optimized the GNR transfer; X-Y.W. and T.D. synthesized the precursor molecules under supervision of A.N. and K.M.; J.O, G.B.B. and M.P. wrote the manuscript; P.R., V.M., R.F. and M.C. supervised the work. All authors commented on the manuscript.

\begin{acknowledgement}

% Please use ``The authors thank \ldots'' rather than ``The
% authors would like to thank \ldots''.

This work was supported by the Swiss National Science Foundation under grant
no. 20PC21\_155644 and the NCCR MARVEL (51NF40\_182892), the European Union’s Horizon 2020 research and innovation program under grant agreement no. 785219 (Graphene Flagship Core 2) and grant agreement no. 767187 (QuIET), and the Office of Naval Research (N00014-18-1-2708 and N00014-12-1-1009).
Part of the computations were performed using resources of the Center for Computational Innovation at Rensselaer Polytechnic Institute.
M.P. acknowledges funding by the EMPAPOSTDOCS-II program, which is financed by the European Union’s Horizon 2020 research and innovation program under the Marie Sklodowska-Curie grant agreement no. 754364.
M.D.L. acknowledges support from the Swiss National Science Foundation Ambizione grant (no. PZ00P2\_179801).
X. Y. W., T. D., A. N., and K. M. acknowledge the support by the Max Planck Society. 
J.O. and O.B. acknowledge technical support from the Binning and Rohrer Nanotechnology Center (BRNC), 
J.O. and M.C. thank R. Furrer, R. Muff, S. Martin, H. Breitenstein for technical support. The authors thank M. El Abbassi, M. Di Giovannantonio, and I. Zardo for fruitful discussions and E. Hack for carefully reading the manuscript.

\end{acknowledgement}

\begin{suppinfo}

% This will usually read something like: ``Experimental procedures and
% characterization data for all new compounds. The class will
% automatically add a sentence pointing to the information on-line:
\begin{itemize}
    \item Sample overview, experimental details and STM data
    \item DFT-calculated mode profiles
    \item Additional Raman data
    \item Damage sensitivity, substrate interaction and length distribution assessment
\end{itemize}

\end{suppinfo}

\bibliography{LCM}

\renewcommand{\thetable}{S\arabic{table}}	% make Supplementary Figures Fig. S XXX
\renewcommand{\thefigure}{S\arabic{figure}} % make Supplementary Figures Fig. S XXX
\setcounter{figure}{0}
\setcounter{table}{0}
\setcounter{table}{0}

\section*{\centering Supplementary Information\\for}
\subsection*{\centering A Universal Length-Dependent Vibrational Mode in\\ Graphene Nanoribbons}

Jan  Overbeck,  Gabriela  Borin  Barin,  Colin  Daniels,  Mickael  L.  Perrin,  Oliver  Braun,  Qiang Sun,  Rimah  Darawish,  Marta  de  Luca, Xiao-Ye Wang,  Tim Dumslaff,  Akimitsu Narita,  Klaus Müllen, Pascal Ruffieux, Vincent Meunier, Roman Fasel, and Michel Calame.\\

\begingroup
\sffamily
Contents:\\
\hyperref[SI: Note 1 - fab and STM]{Supplementary Note 1} - Sample fabrication and STM data\\
\hyperref[SI: Note 2 - Additional Simulations]{Supplementary Note 2} - DFT-calculated mode profiles\\
\hyperref[SI: Note 3 - Additional Raman data]{Supplementary Note 3} - Additional Raman data\\
\hyperref[SI: Note 4 - Damage + Substrate]{Supplementary Note 4} - Damage sensitivity, substrate interaction and length distribution
\endgroup

\vspace{1cm}
\section{\large Supplementary Note 1~-~Sample fabrication and STM data}\label{SI: Note 1 - fab and STM}

Throughout this study, the following types of samples were used: Aligned ribbons grown on vicinal Au crystals -  Au(788) or Au(11,12,12). 
Spectra of AGNRs acquired directly on these Au crystals are referred to as ``on Au(788)''. 
These samples were transferred onto a target substrate using an electrochemical delamination process with a PMMA support layer, referred to as ``bubbling transfer''.\cite{Senkovskiy.2017}
Mostly, these substrates are Raman-optimized (RO) device type substrates.\cite{Overbeck.2019}
Spectra acquired after ``bubbling transfer'' are labelled ``on substrate/BT''.\\
The second type of sample is prepared on thin (200\,nm) gold films supported on Mica (\textit{PHASIS, Geneva,
Switzerland}), which exhibit much wider Au(111) terraces (see Figure~\sfigref[b]{fig:FigSI_Halogens} for an STM image). These samples are referred to as ``on Au/Mica''.
They are transferred onto substrates in a polymer-free transfer by cleaving off the mica and etching away the Au-layer.\cite{BorinBarin.2019} 
Spectra acquired after the ``polymer-free transfer'' method are labelled ``on substrate/PFT''.\\
Finally, AGNRs of the same width can be produced from different precursor molecules, specifically, precursors with reactive groups based on bromine or, alternatively, iodine.
These details are listed in Table~\ref{tab:SI_sample_list} below.\\

\renewcommand{\arraystretch}{1.5}
\begin{table}[htb]
    \centering
    \begin{tabular}{l|c|c|c|c}
        Sample label & Precursor-halogen & Substrate & Transfer method & Figures  \\ % our old label:
        \hline
        5-1\_Au     & Iodine & Au(788) & - & \sfigref[a, b]{fig:Fig1},  \sfigref[c]{fig:FigPol2}, \sfigref[b]{fig:FigSI_STM_bare_Au788_5.I.01} \\ % 5.I.01
        5-2\_Au     & Iodine & Au/Mica & - & \sfigref[c]{fig:FigTheo3}, \sfigref[a]{fig:FigSI_Halogens}, \sfigref[a]{fig:FigSI_5-before_after} \\ %  5.I.06,
        5-3\_Au     & Iodine & Au(788) & - & \sfigref[c]{fig:FigTheo3}, \sfigref[a]{fig:FigSI_Halogens} \\ %  5.I.08,
        5-4\_Au     & Iodine & Au/Mica & - & \sfigref[c]{fig:FigTheo3} \\ %  5.I.07, 
        5-5\_Au     & Iodine & Au/Mica & - & \sfigref[c]{fig:FigTheo3}, \sfigref[a]{fig:FigSI_Halogens} \\ %  5.I.04
        5-6\_Au     & Bromine & Au(788) & - & \sfigref[b]{fig:FigTheo3}, \sfigref[a]{fig:FigSI_Halogens}, \sfigref[b]{fig:FigSI_5-before_after} \\ % 5.Br.02
        5-6\_T     & Bromine & RO & BT & \sfigref[b]{fig:FigSI_5-before_after} \\ % 5.Br.02b
        5-7\_T       & Iodine & Si/SiO\tsub{2} & BT & \sfigref[a]{fig:FigSI_5-before_after} \\ % 5.I.05
        5-8\_Au      & Bromine & Au(788) & - &
        \ref{fig:FigSI_STM_LT-short5},
        \sfigref[a]{fig:FigSI_Halogens}, \ref{fig:FigSI_PolDep_short5} \\ % 5.Br.06
        5-8\_T       & Bromine & RO & BT & \sfigref[b]{fig:FigSI_PolDep_short5} \\ % 5.Br.06 
        5-9\_Au       & Iodine & Au/Mica & - & \sfigref[a, b]{fig:FigSI_Halogens} \\ % 5.I.11
        5-10\_Au    & Bromine & Au/Mica & - & \sfigref[a]{fig:FigSI_Halogens} \\ % 5.Br.03

        \hline
        % 7-1\_Au     & Bromine & Au/Mica & - & Fig? \\ % 7.Br.01
        7-1\_T     & Bromine & RO & PFT & \sfigref[b]{fig:Fig1}, \sfigref{fig:FigSI_BG_sub}\\ % 7.Br.01 transferred
        7-2\_T     & Bromine & RO & BT & \sfigref[a]{fig:FigSI_PolDep_7-9} \\ % 7.Br.02
        \hline
        9-1\_T     & Iodine & RO & BT & \sfigref[b]{fig:Fig1}, \sfigref[b]{fig:FigSI_Substrate-Interaction} \\ % 9.I.06 transferred
        9-2\_T     & Iodine & RO & PFT & \sfigref[a]{fig:Fig4}, \sfigref{fig:FigSI_Damage_full}\\ % 9.I.07 transferred
        9-3\_Au     & Iodine & Au(788) & - & \sfigref[a]{fig:FigSI_extra9} \\ % 9.I.03 
        % 9-4\_T      & Bromine & Au(11,12,12) & - & \sfigref[b]{fig:FigSI_extra9} \\ % 9.Br.01  ---mistakenly thought this was the WL-dep
        9-4\_T      & Iodine & Au(11,12,12) & - & \sfigref[b]{fig:FigSI_extra9}, \sfigref[b]{fig:FigSI_PolDep_7-9} \\ % 9.I.01
    \end{tabular}
    \caption{Details for samples used throughout this study.}
    \label{tab:SI_sample_list}
\end{table}
\renewcommand{\arraystretch}{1}

\newpage

\subsection*{\normalsize STM images of GNRs on growth substrates}\label{SI: 1.1 STM}

STM imaging is the prime technique for investigating GNRs on their growth substrate in UHV, occasionally complemented with non-contact atomic force microscopy.\cite{Cai.2010} 
Here we focus on two types of STM, room-temperature STM (RT-STM or just STM) and low-temperature STM (LT-STM).
Figure~\sfigref[a]{fig:FigSI_STM_bare_Au788_5.I.01} below shows STM images of a bare Au(788) vicinal crystal, showing the precisely aligned terraces.
Figure~\sfigref[b]{fig:FigSI_STM_bare_Au788_5.I.01} displays a Au(788) crystal after growth of 5-AGNRs. The GNRs grow predominantly along the direction of the terraces (see inset). In contrast to 7- and 9-AGNRs, significant step bunching is observed for 5-AGNRs, resulting in wider terraces compared to the bare substrate. 
Here, the 5-AGNRs are visible for imaging at room temperature, because the sample contains predominantly long GNRs, which are relatively immobile.
The Raman spectrum in Figure~\ref{fig:Fig1} of the main manuscript was acquired on this particular sample (5-1\_Au).

\begin{figure}[ht!]
    \centering
    \includegraphics[width=\textwidth]{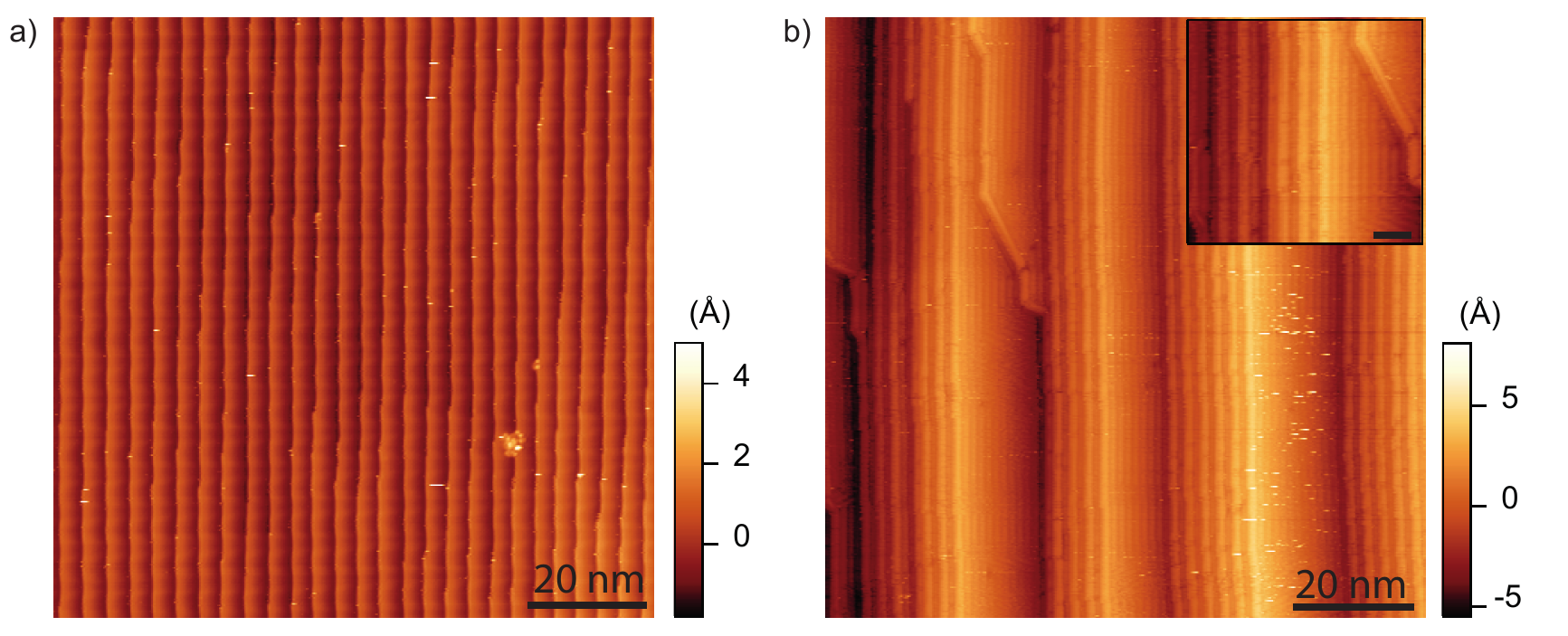}
    \caption{\textbf{Room-temperature STM of a Au(788) growth substrate before and after growth of long 5-AGNRs. a)} Bare Au(111) terraces on a Au(788) single crystal substrate. \textbf{b)} Long, aligned 5-AGNRs grown on Au(788) with significant step-bunching. Inset: zoom onto one step revealing parallel GNRs. Scale bar: 5\,nm. Sample~5-1\_Au}
    \label{fig:FigSI_STM_bare_Au788_5.I.01}
\end{figure}

\subsubsection*{LT-STM on short 5-AGNRs}

\begin{figure}[ht]
    \centering
    \includegraphics[width=\textwidth]{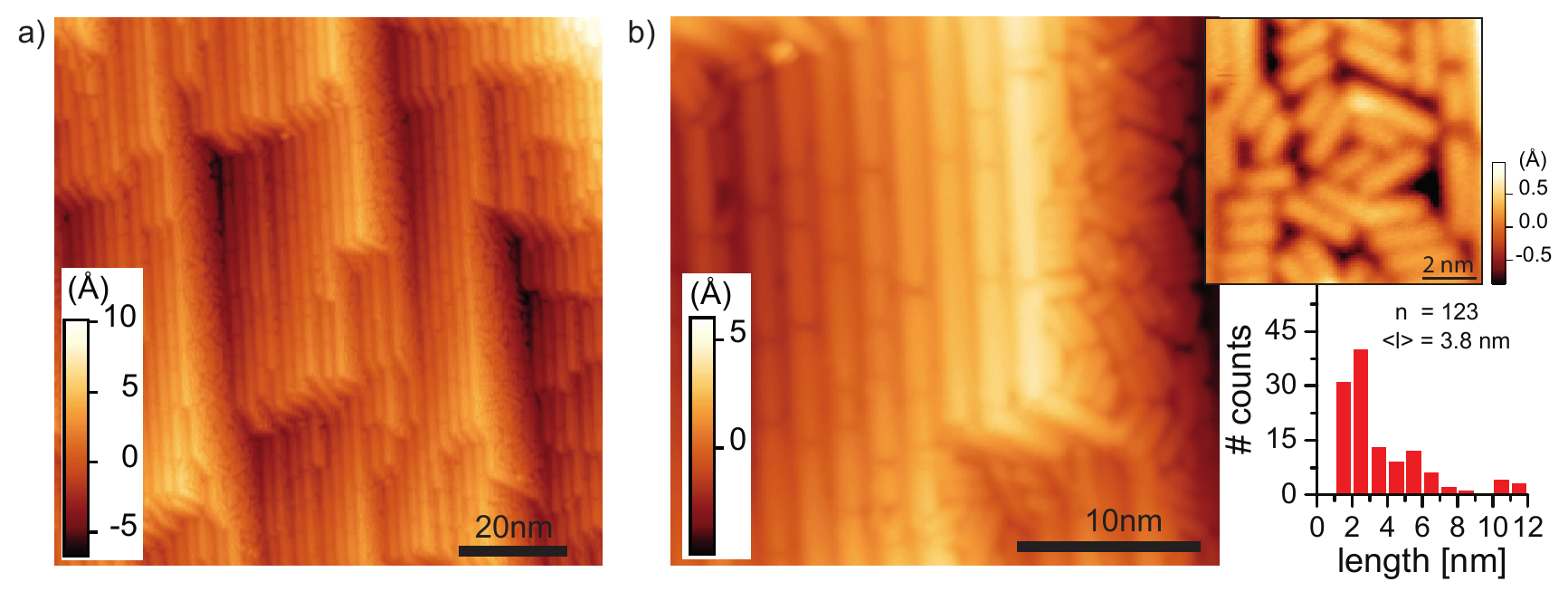}
    \caption{\textbf{Low temperature STM images of short, aligned 5-AGNRs. a)} Overview scan of a sample produced from partially debrominated precursor molecules resulting in short 5-AGNRs on a Au(788) crystal with pronounced step-bunching. \textbf{b)} High resolution scan from a) clearly revealing a large amount ($\sim 50\%$, see histogram) of 4-, 6-\&~8-unit 5-AGNRs. Sample~5-8\_Au.} % 5.Br.06.
    \label{fig:FigSI_STM_LT-short5}
\end{figure}

Samples containing short ribbons are difficult to image at room temperatures, because smaller GNRs are dragged along the surface by the scanning probe, resulting in fuzzy areas on the image. 
For these samples, high-resolution LT-STM (T=5\,K) images were recorded to visualize these short 5-AGNRs. Figure~\ref{fig:FigSI_STM_LT-short5} shows two images of a sample that contains a large number of very short GNRs. 
The corresponding Raman spectrum is shown in Figure~\sfigref[a]{fig:FigSI_Halogens}, labelled 5.Br.06.
The inset of Figure~\sfigref[b]{fig:FigSI_STM_LT-short5} clearly shows an assembly of very short (2-5\,nm) GNRs. Note, that the short ribbons on the larger terraces are nearly randomly oriented, while they are aligned with the crystal direction on the narrower terraces.
An image processing script was used to extract the length of GNRs in well-resolved areas. The corresponding histogram for this sample is shown as an inset, confirming a majority of very short GNRs. Specifically, the GNRs with a length below 2/3/4\,nm correspond to 4/6/8-unit 5-AGNRs (see Figure~\ref{fig:FigSI_ModeProfiles} for corresponding structures).

\subsubsection*{Correlation of Raman spectra with STM images.}\label{SI: Halogens}

\begin{figure}[t!]
\centering
\includegraphics[width=\textwidth]{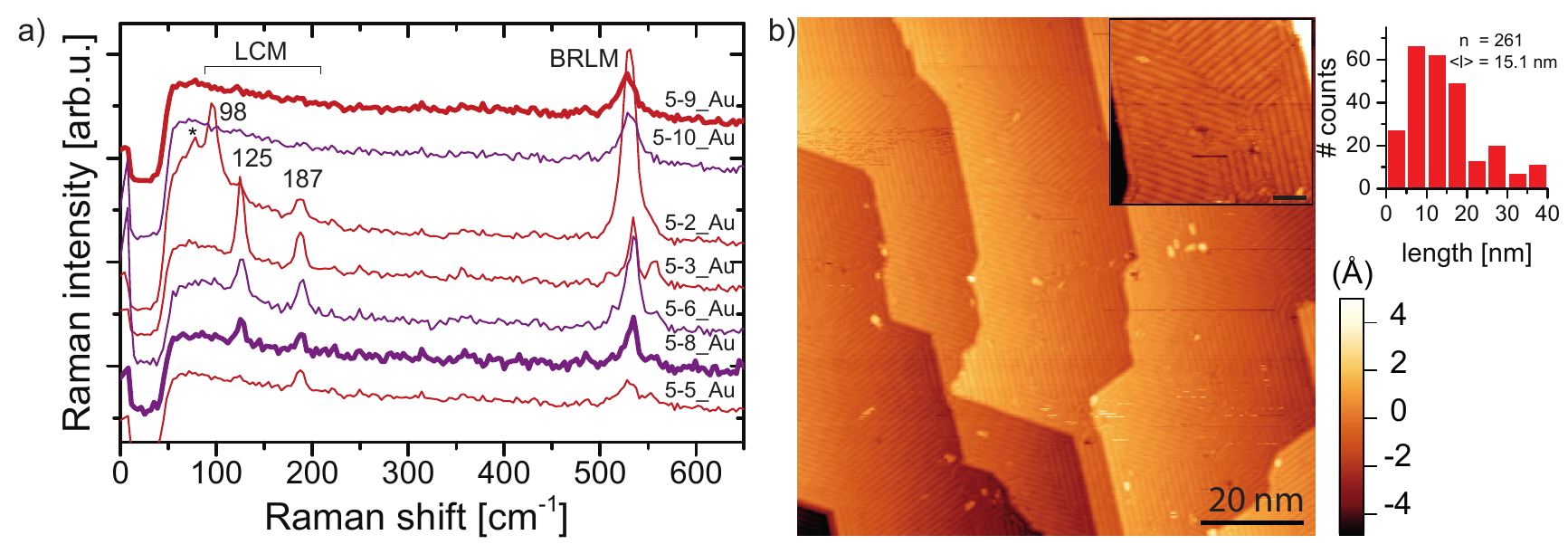}
\caption{\textbf{Sample comparison of 5-AGNRs on gold with different LCM-contributions for both precursor halogens. a)} Raman spectra for 5-AGNR samples with different length distributions. Samples synthesized from both Br/I-based precursors (purple/red) are shown to produce both long and short GNRs. Spectra shown with bold lines correspond to STM images in panel b) (Sample~5-9\_Au) and Figure~\ref{fig:FigSI_STM_LT-short5} (Sample~5-8\_Au). % 5-9, 
\textbf{b)} STM image of Sample~5-9\_Au containing predominantly long 5-AGNRs. Scale bar of inset 5\,nm. Histogram shows the length distribution extracted from the STM image.} % 5.I.11
\label{fig:FigSI_Halogens}
\end{figure}

For the case of 5-AGNRs, in total, we investigated 15~samples (9~shown here) directly on the growth substrates, Au(788) and Au/Mica. Of these, 12 (3) were grown from iodine(bromine)-based precursors, respectively. 
A study about the optimization of GNR growth parameters will be published elsewhere.
The STM characterization of most samples was carried out at room temperature. Therefore, the amount of short GNRs is inferred from the prevalence of fuzzy areas/ areas not showing long, well-aligned AGNRs.
In Figure~\sfigref[a]{fig:FigSI_Halogens} we plot Raman spectra of 7 representative samples. All samples, except those with optimized growth recipes, show the LCM attributed to 4-/6-/8-unit long 5-AGNRs. 
The STM image of an optimized sample grown on Au/Mica is shown in Figure~\sfigref[b]{fig:FigSI_Halogens}, along with a histogram of ribbon length. Note, that very long GNRs extending beyond the scan range were not counted into the histogram at all which likely leads to an underestimated average GNR length.
Nevertheless, with an optimized growth recipe one achieves lengths comparable to that of previously optimized 7- and 9-AGNR growth recipes.\cite{DiGiovannantonio.2018, BorinBarin.2019} 

For the samples with a significant amount of short ribbons we identify the characteristic LCM peak positions at Raman shifts of approximately 98\tcm,~ 125\tcm~and 186\tcm~as well as the RBLM at about 530\tcm.
The spectra are arranged according to the presence of the LCM, with increasing contribution of LCM-98/125/186 from top to bottom.
We do not observe any change in frequency of these peaks when changing between bromine- and iodine-based precursors.
The trend observed here agrees well with the observations from RT-STM.
A quantitative analysis is hampered by the prohibitive time consumption for low-temperature STM analysis on each sample, highlighting the benefit of rapid identification of short GNRs by Raman spectroscopy.

\newpage
\section{\large Supplementary Note 2~-~DFT-calculated mode profiles}\label{SI: Note 2 - Additional Simulations}

\begin{figure}[!htb]
\centering
\includegraphics[width=\textwidth]{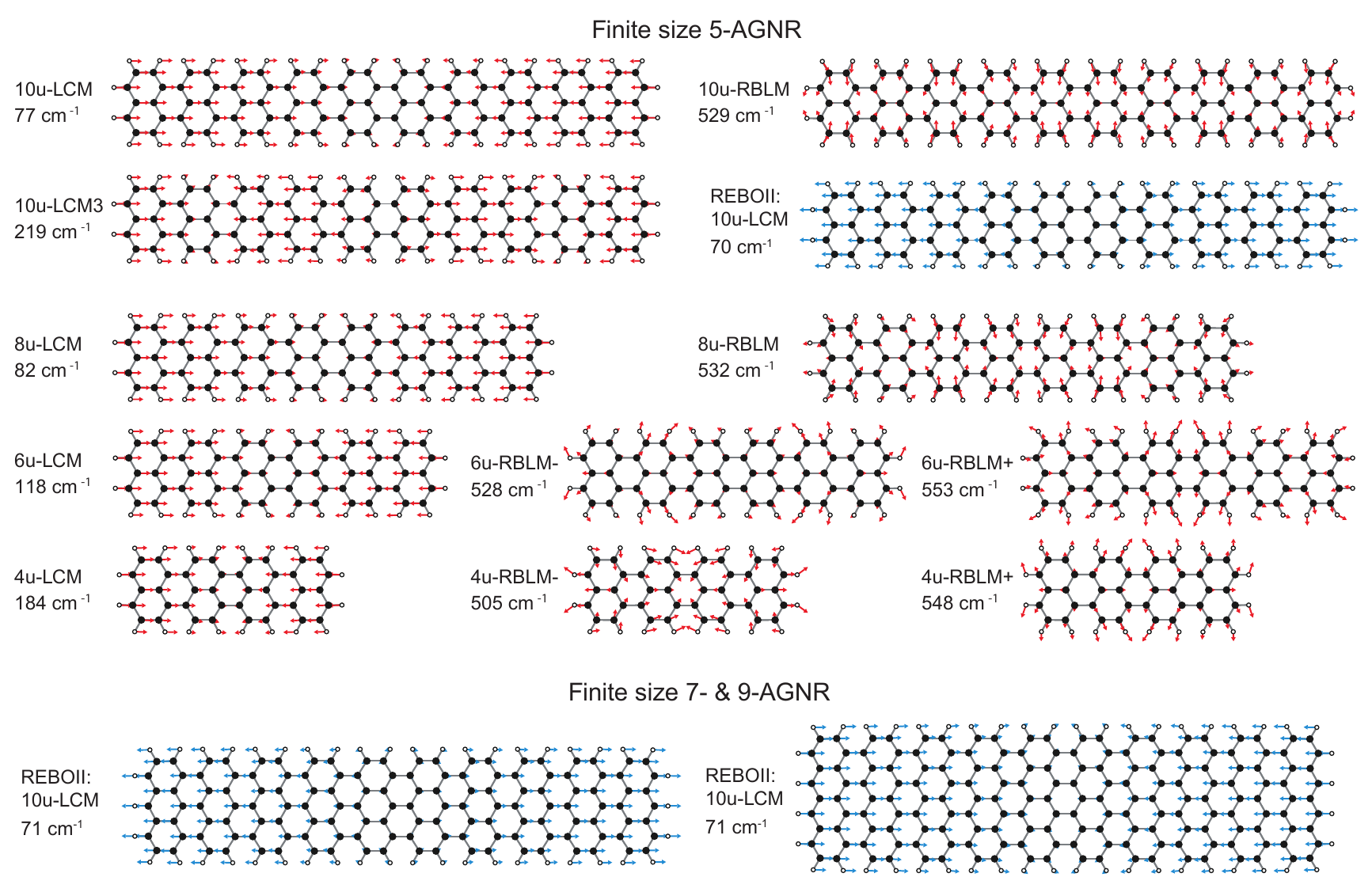}
\caption{\textbf{Normal mode profiles obtained from DFT/REBOII calculations (red/blue).}}
\label{fig:FigSI_ModeProfiles}
\end{figure}

In addition to the mode frequencies and Raman intensities we also obtained the normal mode profiles from DFT- and REBOII-based calculations.
Figure\,\ref{fig:FigSI_ModeProfiles} shows the atomic displacement profiles from DFT for the LCM and its overtone, LCM3, for a finite size 10\,units long 5-AGNR (10u-LCM etc.), as well as the normal mode profile for the RBLM.
In addition, we plot the LCM and RBLM for the 8\,unit, 6\,unit, and 4\,unit 5-AGNRs.
Note, that for the latter two, the normal mode profile for the RBLM looses its purely transverse character as a result of the finite length of the GNR.
We plot the profiles for both RBLM+ and RBLM- observed in the Raman spectra of samples with many short GNRs.
We also show the LCM computed with the REBOII force-field for the three GNR widths of 10\,units length (blue).
We observe a lower frequency compared to the DFT-based calculations (5-AGNR) and experiments (7-/9-AGNR).
For the other normal modes of the 7- and 9-AGNR we refer to our previous work.\cite{BorinBarin.2019}

\newpage
\section{\large Supplementary Note 3~-~Additional Raman data}\label{SI: Note 3 - Additional Raman data}

\subsection*{\normalsize Measurement considerations}\label{SI: 3.1 Raman Methods}

Acquisition of high quality Raman spectra relies on the optimization of several measurement parameters that have to be adjusted to the GNR under investigation.
In general, we follow a mapping strategy and rely on Raman optimized (RO) device-type substrates (consisting of a layer structure of  Si/SiO\tsub{2} and Au/Al\tsub{2}O\tsub{3}), as we have recently described elsewhere.\cite{Overbeck.2019} \\
Here we highlight a few points of particular importance:

\subsubsection*{Excitation wavelength}
We mainly used three wavelengths, 488\,nm, 532\,nm and 785\,nm to characterize the samples (a further resonance Raman study combining measurements on multiple setups confirm these data and will be published elsewhere). For each type of GNR the excitation wavelength was chosen to give the strongest possible signal for the LCM.

The 5-AGNR has a small band gap in the range of $\sim$\,0.5\,eV, as deduced from STS data, and $\sim$\,1.7\,eV predicted by GW calculations before taking exciton binding into account\cite{Yang.2007}. 
Therefore, we predominantly used 785\,nm (1.57\,eV) for excitation, which is the lowest laser energy available in our setup and thus closest to resonance with the fundamental band gap.
For the 7-AGNR, the 532\,nm (2.33\,eV) laser lies in between the reported $\text{E}_{1,1} \text{ and E}_{2,2}$ transitions and therefore gives a resonantly enhanced Raman signal. \cite{Senkovskiy.2017, BorinBarin.2019} 
Lastly, GW-calculations predict a band gap of 2.1\,eV for the 9-AGNR, with strong exciton binding resulting in a lowest absorption peak expected at about 1\,eV.\cite{Prezzi.2008, Zhao.2017} 
Even though the 785\,nm laser is closer to the lowest energy transitions and results in the best signal for the RBLM, we observe the highest intensity of the LCM at an excitation wavelength of 488\,nm. This points towards an electronic transition at a higher energy which is more strongly coupled to this particular vibrational mode. %\todo{Read and find more info from e.g. https://pubs.acs.org/doi/pdf/10.1021/j100161a077}.

\subsubsection*{Mapping, background subtraction and fitting}

Throughout this work we employ a mapping strategy to acquire of high signal-to-noise spectra by averaging (up to several 1000s of spectra), to make sure we report the average GNR properties and to exclude outliers produced by local contaminants.
This mapping approach also simplifies the identification of substrate effects.
The inset in Figure~\sfigref[a]{fig:FigSI_BG_sub} shows the outline (red line) of such a typical map on an RO device substrate. 
The edge of a GNR film is highlighted by a dashed white line. 
We average over a homogeneous part of such a Raman map (e.g. the bottom left quarter of that map) to get a high signal-to-noise spectrum.
Such a map at the edge of the GNR film can further be used, to identify and exclude spurious effects from the substrate or setup.

\begin{figure}[!htb]
\centering
\includegraphics[width=\textwidth]{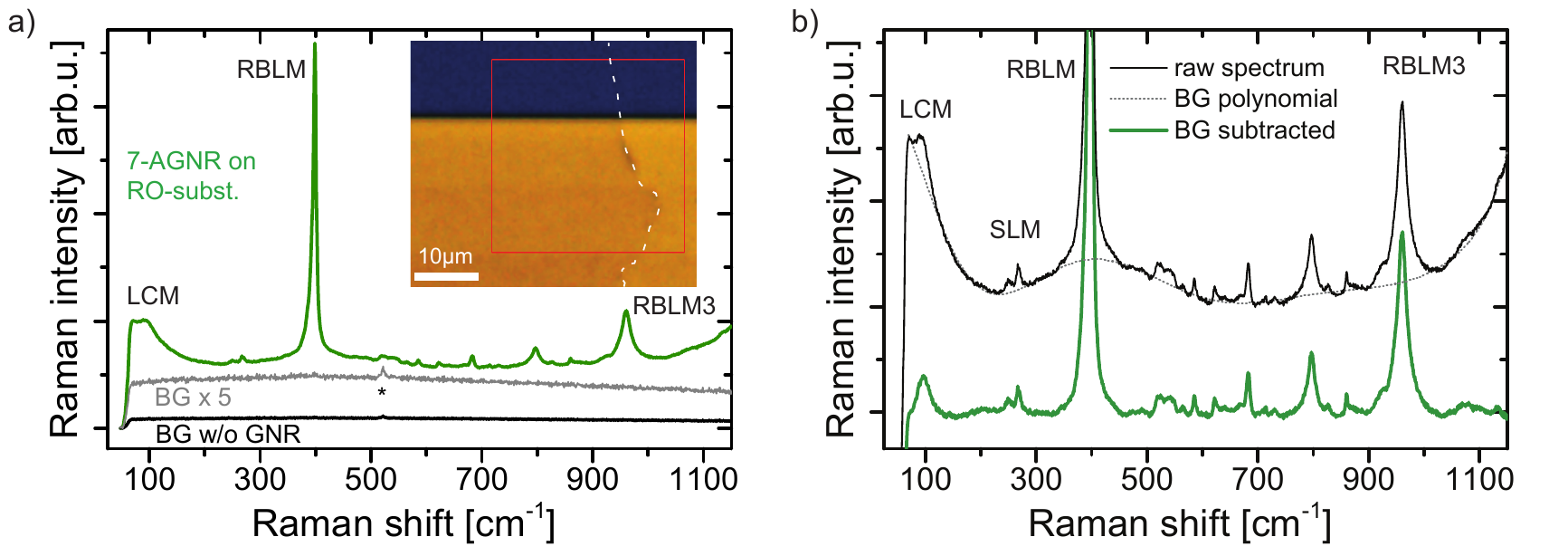} % Sample 7.Br.01
\caption{\textbf{Mapping and background subtraction from raw data exemplified on 7-AGNRs. a)} Raw spectra of 7-AGNRs grown on Au/Mica transferred to an RO-substrate. The inset shows an optical image of the sample area with the scan area indicated by a red square. The edge of the GNR film can be identified from the change in contrast on the Au-pad of the RO-substrate and is highlighted by a white dashed line. The GNR spectrum is obtained by averaging spectra from the bottom left side of the scan, the background spectrum by averaging in the GNR-free area of the same exact scan. \lex = 532\,nm, average over 100 (22) pixels for the GNR (BG) spectrum. Sample~7-1\_T.
\textbf{b)} Zoom in to the raw spectrum from a) before and after polynomial background subtraction. The polynomial is indicated by a dotted line.}
\label{fig:FigSI_BG_sub}
\end{figure}

\begin{figure}[!htb]
\centering
\includegraphics[width=\textwidth]{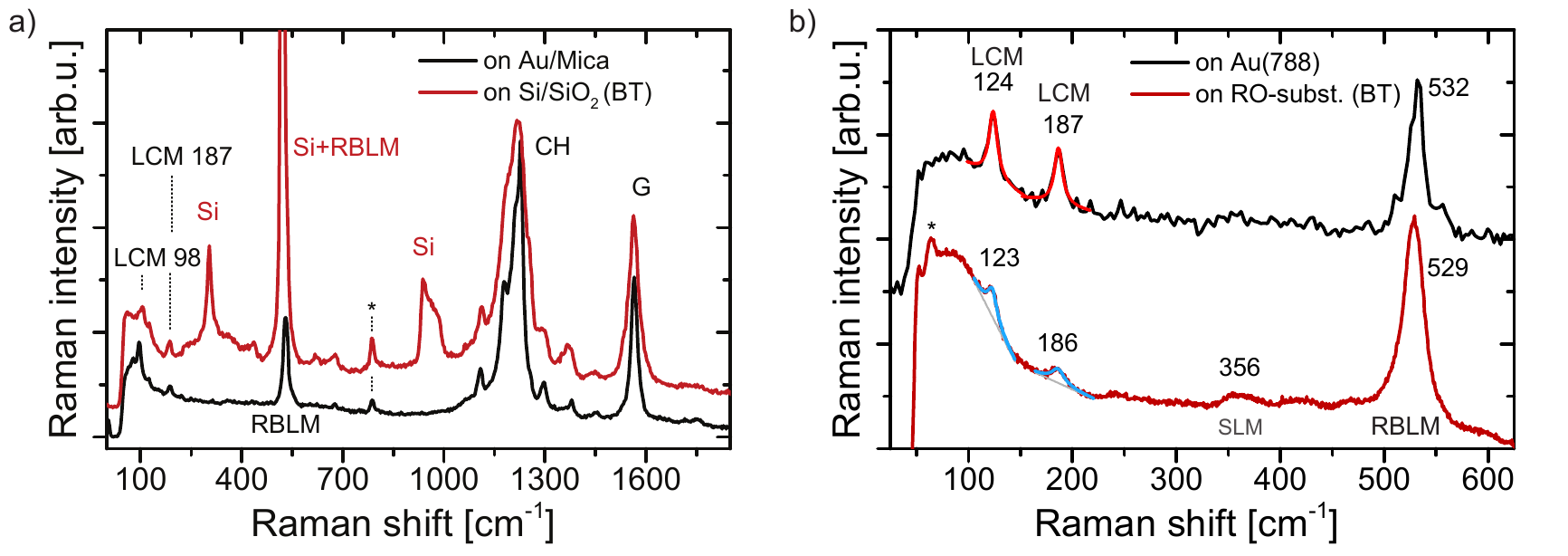}
\caption{\textbf{Spectra of 5-AGNRs on gold and after transfer to substrates. a)} Full spectral range raw spectrum of 5-AGNRs on Au/Mica including background. This particular sample (5-2\_Au) could not be imaged after transfer. A sample grown one day before with the same parameters but on Au(788) is shown after transfer to a standard Si-substrate. The most prominent peaks are labelled and peak positions are indicated. The peak labelled with an asterisk is discussed elsewhere.\cite{Overbeck.2019} $\lambda_{exc}$ = 785\,nm, Sample~5-7\_T. 
% Sample 5.I.06- AuMica before transfer and 5.I.05-Au788 after transfer! 
\textbf{b)} Raman spectra of short 5-AGNRs grown on Au(788) before and after ``bubbling transfer''. The most prominent peaks are labelled and peak positions as obtained from Lorentzian fits are indicated. The intensity oscillations at and below $\sim 75\,$\tcm~marked with an asterisk are artefacts from the 785\,nm laser edge filter. \lex = 785\,nm, Samples 5-6\_Au, 5-6\_T.} % Sample 9.Br.02, 9.Br.02b
\label{fig:FigSI_5-before_after}
\end{figure}

\begin{figure}[!htb]
\centering
\includegraphics[width=\textwidth]{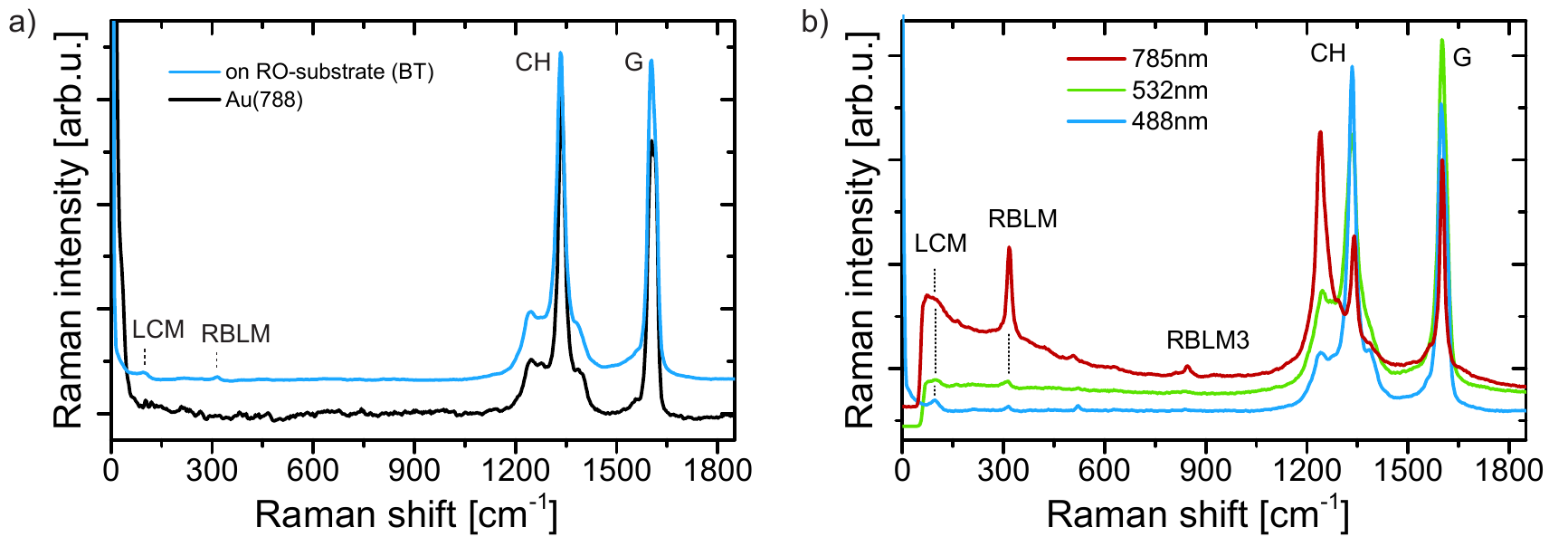}
\caption{\textbf{Raman spectra of 9-AGNRs. a)} Comparison of a spectrum on Au(788), 9-3\_Au, on an RO-substrate, Sample~9-1\_T, scaled to a common G-peak intensity. $\lambda_{exc}$ = 488\,nm.  %Samples 9.I.03 before and 9.I.06 after
\textbf{b)} Raman spectra of 9-AGNRs after transfer to an RO-substrate for three excitation wavelengths. These spectra were acquired under ambient conditions and are arbitrarily scaled to highlight low-intensity features. Sample~9-4\_T.} % Sample 9.I.01
\label{fig:FigSI_extra9}
\end{figure}

Depending on the substrate and excitation wavelength, the raw spectra contain more or less background originating from both GNRs and the substrate.\cite{Senkovskiy.2017} 
To facilitate comparison between spectra obtained on different substrates and with different lasers (more or less resonant with the particular GNRs), we subtract a polynomial background fitted to the raw spectrum (masking peaks from the fit). 
Figure~\ref{fig:FigSI_BG_sub} shows an example of the procedure for a sample of 7-AGNRs transferred from Au/Mica.

Spectra for GNRs and background are obtained by averaging over the respective areas of the scan. The background obtained on the bare substrate after the Au-etching procedure is relatively low and completely featureless, except for a weak Raman peak of the Si-substrate (indicated by an asterisk) which originates from the residual transparency of the gold film. 
In particular, there is no (substrate- or setup-induced) intensity modulation in the background spectrum at low wavenumbers close to the onset of the filter.
In contrast, the Raman spectrum shows several well defined peaks.
Figure~\ref{fig:FigSI_BG_sub}\,b) shows the raw spectrum on a zoomed in scale, as well as the polynomial used for background subtraction and the resulting spectrum. 
Several settings for fitting the background polynomial were explored resulting in minor variability of absolute peak intensity and negligible influence on peak positions.
Peak positions were obtained by fitting with a Lorentzian or Voigt profile.

\subsubsection*{Laser power}

The GNRs are in contact with two fundamentally different substrates during Raman investigation, the Au substrate and a dielectric (SiO\tsub{2} or Al\tsub{2}O\tsub{3}).
Similar to graphene, where it is well-known that the substrate strongly influences the optical properties, we also observe a much stronger Raman signal of the GNRs after transfer to a dielectric. 
Therefore, in analogy to Figures\,\ref{fig:Fig4} and \ref{fig:FigSI_Damage_full}, we performed time-series measurements for each of the substrate-laser-wavelength combinations to choose the optimum integration-time and laser power while avoiding radiation-induced damage to the GNRs.

\subsection*{\normalsize Raman spectra before \& after transfer}\label{SI: Add_Raman}

Below we show spectra of 5-AGNRs both on growth substrate and after transfer. The fact that we can observe the LCM-187/125/98 after transfer clearly indicates the presence of short GNRs on the target substrate which can so far not be detected by other means. The slightly reduced intensity with respect to the RBLM fits to a lower transfer efficiency of short vs. long GNRs.
For the LCM-98 we observe a slight up-shift after transfer to an oxide substrate, in line with our initial substrate-dependent findings on 9-AGNR as described in the main manuscript.

\newpage

\subsection*{\normalsize Spectra of 9-AGNRs on gold \& on device-substrates and initial wavelength dependence}\label{SI: Wavelength}

In Figure~\sfigref[a]{fig:FigSI_extra9} we compare the Raman spectrum of 9-AGNRs obtained directly on Au(788) and after transfer to an RO-substrate. 
The much lower intensity of GNR Raman signal on gold and the background from the metallic Au(788) substrate prevent us from detecting or excluding the presence of the LCM on the growth substrate.

Figure~\sfigref[b]{fig:FigSI_extra9} shows the spectrum of 9-AGNRs after BT onto an RO-substrate for three laser wavelengths.
The LCM is most clearly observed at an excitation wavelength of \lex = 488\,nm which corresponds to an energy significantly above the expected band gap of the 9-AGNR. For the other wavelengths, the LCM is less clear due to the onset of the edge filter in our setup. More extensive wavelength-dependent measurements are beyond the scope of this paper and are the subject of an ongoing study.

\subsection*{\normalsize Polarization dependent measurements}\label{SI: PolarizationExpl.}

We also performed polarization dependent measurements on aligned 7- and 9-AGNRs before (not shown) and after transfer. 
Unsurprisingly, these GNRs also show a strong dependence of Raman intensity on the Laser polarization direction $\theta$. We fit each peak with a Lorentzian line shape and a linear background. 
The peak intensity follows a broadened cos\textsuperscript{2}($\theta$) dependence as  expected for a collection of dipoles (for an experimental configuration averaging over all angles in the detection path).\cite{Senkovskiy.2017} 
We use a simplified fitting model
\begin{equation}
\centering
    I(\theta) =  a\cdot\text{cos}^2(\theta-\theta_0)+b\cdot\text{sin}^2(\theta-\theta_0),
\end{equation}
where $\theta_0$ is the misalignment between the Au(788)-terrace direction and the experimental axis defined as $\theta = 0$.

From this we extract the polarization anisotropy \begin{equation}
\centering
    P = \frac{I_\parallel - I_\perp}{I_\parallel + I_\perp} = \frac{a - b}{a + b}.
\end{equation}

\begin{figure}[ht]
    \centering
    \includegraphics[width=\textwidth]{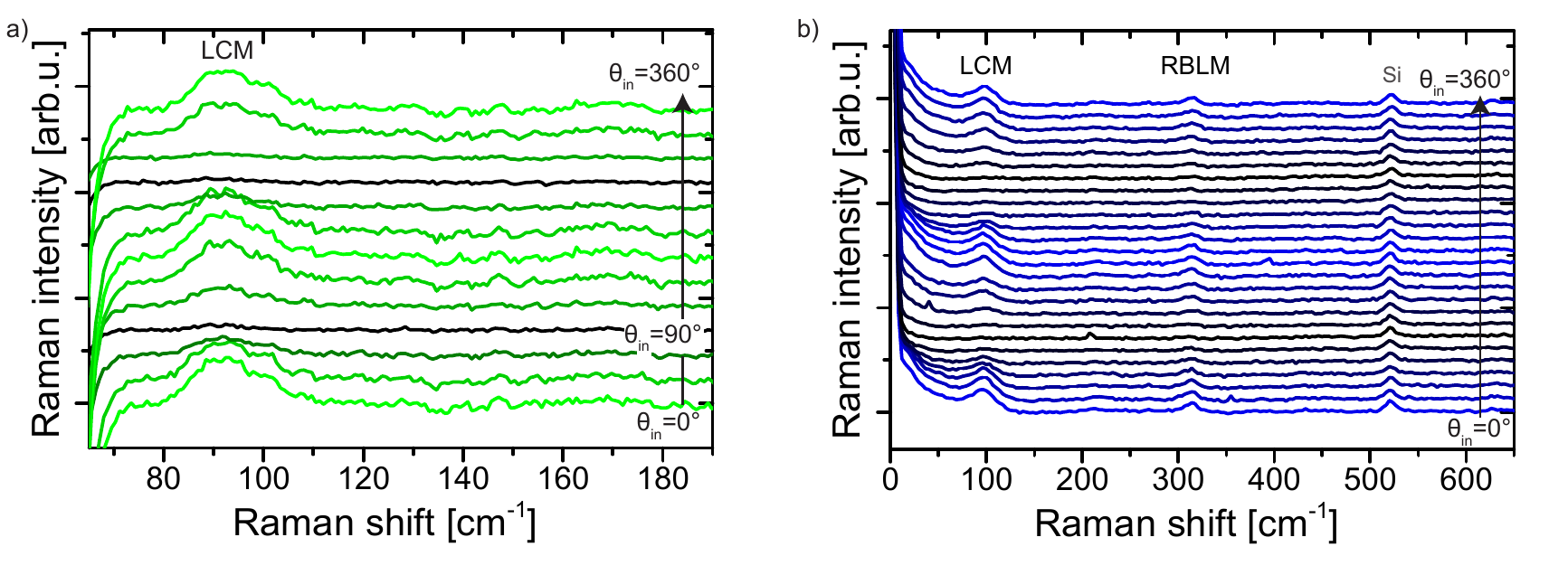}
    % \missingfigure[figwidth=17.5cm, figheight=6cm]{Raman Poldep 7- + 9-AGNR}
    \caption{\textbf{ Polarization dependent spectra of the LCM for aligned 7- and 9-AGNRs. a)} Polarization dependence of the LCM on a sample of aligned 7-AGNRs transferred to an RO-substrate. The polarization of the excitation was changed in steps of $\SI{30}{\degree}$. No polarizer was used in the detection path to maximize the overall signal. Large area average over 400 pixels. $\lambda_{exc}$ = 532\,nm, Sample~7-2\_T. % Sample 7.Br.02
    \textbf{b)} Polarization dependence for aligned 9-AGNR on an RO-substrate. The polarization of the excitation was changed in steps of $\SI{15}{\degree}$. No polarizer was used in the detection path to maximize the overall signal. This measurement was performed under ambient conditions at an excitation wavelength of \lex= 488\,nm. Each spectrum is an average over 1430 pixels of a large area scan with reduced integration time, to avoid laser-induced damage to the ribbons. Sample~9-4\_T.} % Sample 9.I.01
    \label{fig:FigSI_PolDep_7-9}
\end{figure}

Figure~\ref{fig:FigSI_PolDep_7-9} shows the polarization dependence of the low-energy region of the spectra for these two types of ribbons. 7-AGNRs were measured with resonant 532\,nm excitation in vacuum to get the best signal-to-noise ratio.
The LCM of 9-AGNRs is most clearly observed for an excitation wavelength of 488\,nm, where our setup allows experiments at low Raman shifts. 
We measured consistent spectra of 9-AGNRs both in vacuum (not shown) and in air, which allows us to exclude any contribution from the window of the vacuum chamber.

We determine the polarization anisotropy for the different peaks. For the LCM, RBLM and G-peak (not shown) we find $\text{P\tsub{LCM}} = 0.97$, $\text{P\tsub{RBLM}} = 0.73$, $\text{P\tsub{G}} = 0.72$.

\begin{figure}[ht]
    \centering
    \includegraphics[width=\textwidth]{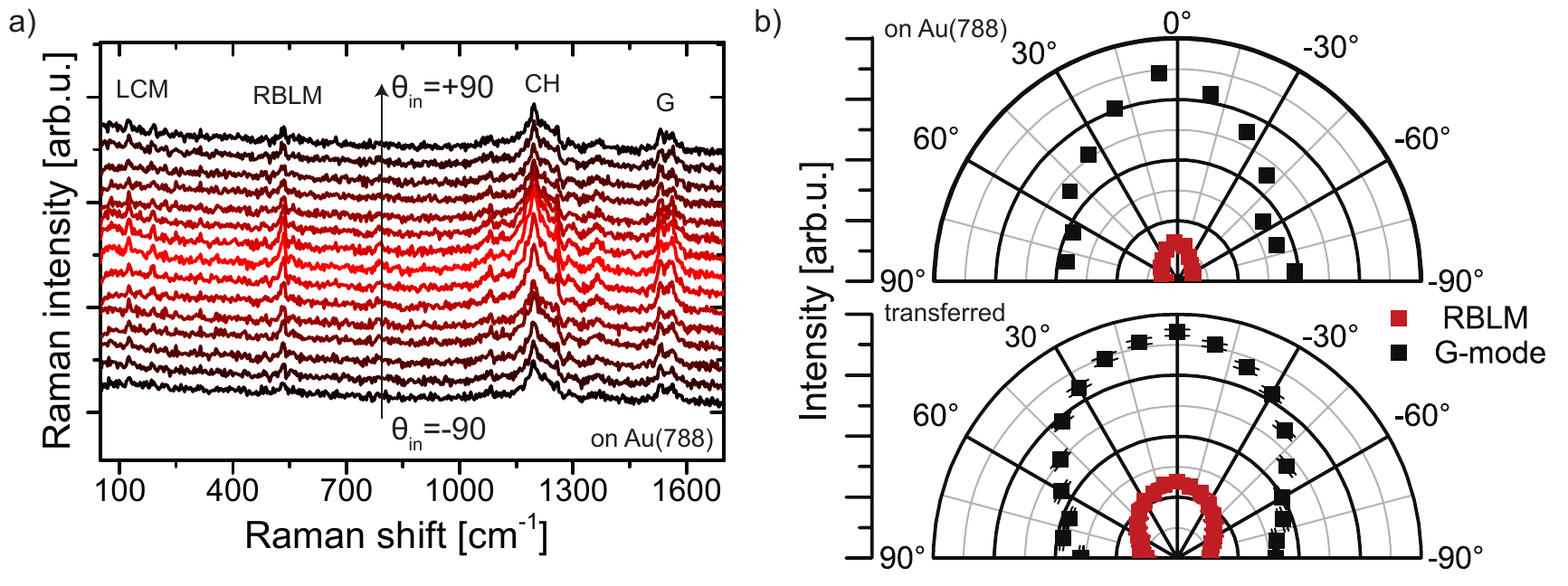} % Sample 5.Br.06
    \caption{\textbf{Polarization dependent spectra of short 5-AGNRs grown on Au(788) a)} Raman spectra of short 5-AGNRs obtained directly on the Au(788) growth substrate. Analyzing polarizer\,$\parallel\theta_{in}$. \lex = 785\,nm. Sample~5-8\_Au.
    \textbf{b)} Polar plots for RBLM and G-peak before and after GNR transfer (Sample~5-8\_T). The polarization anisotropy is $\text{P\tsub{RBLM, (G)}} = 0.4, (0.3)$ before and $\text{P\tsub{RBLM, (G)}} = 0.4, (0.4)$ after transfer (no analyzing polarizer).}
    \label{fig:FigSI_PolDep_short5}
\end{figure}

\begin{figure}[!htb]
    \centering
    \includegraphics[width=\textwidth]{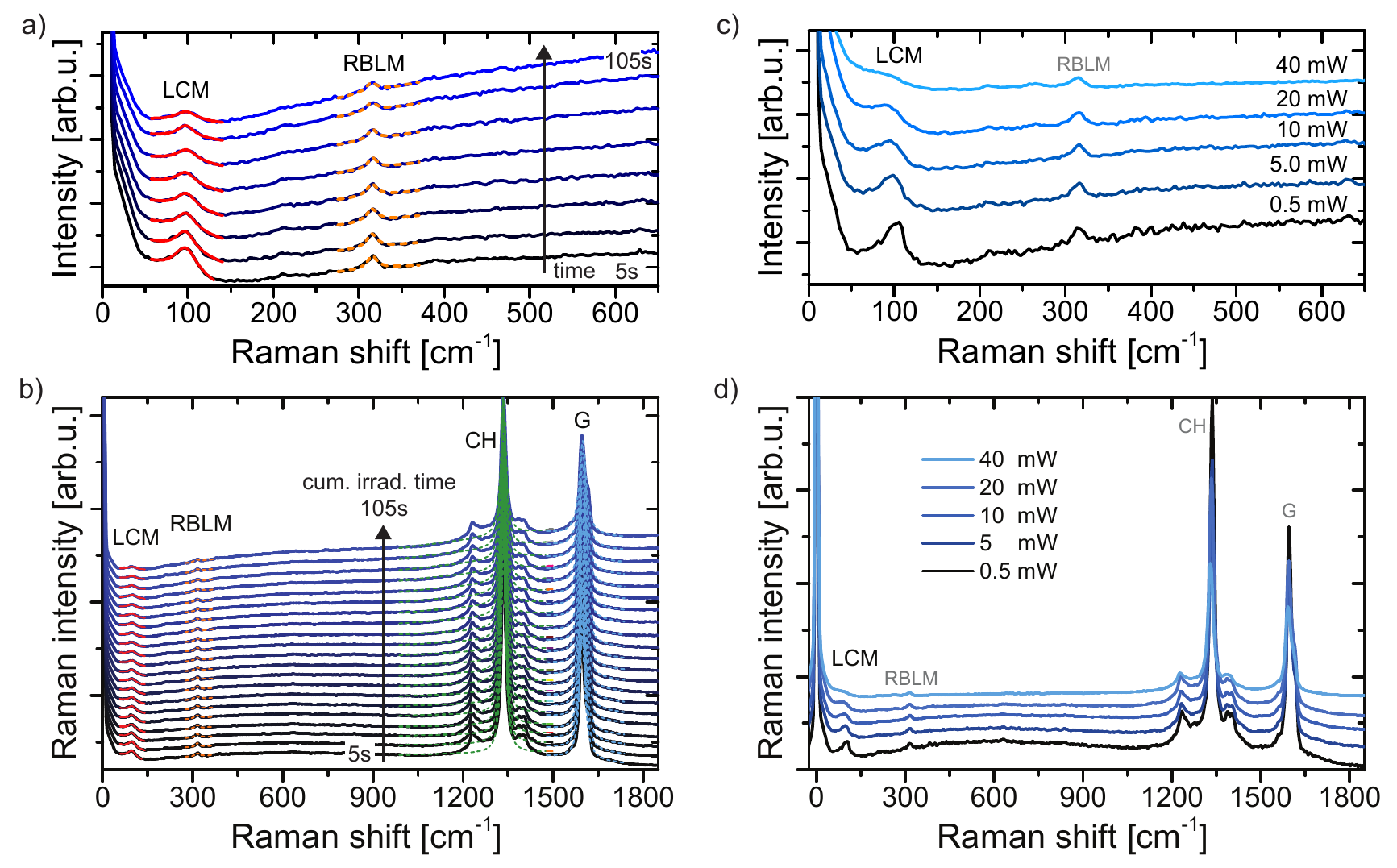} % Sample 9.I.07
    \caption{\textbf{Radiation damaging of 9-AGNRs. a)} Low-frequency Raman spectra over time, $t_{\text{int}} = \SI{5}{\second}$, 15\,s apart, $P=\SI{5}{\milli\watt}$. LCM and RBLM are fitted with a Lorentzian and linear background. \textbf{b)} Full spectral range of a) with the full series of spectra 5\,s apart, including fits to CH- and G-peak.
    \textbf{c)} Power-dependent damaging. Spectra are scaled to a constant power-integration time product (0.5\,mW\,x\,1.1\,s). \textbf{d)} Corresponding full-range spectra. All Sample~9-2\_T and \lex = 488\,nm.}
    \label{fig:FigSI_Damage_full}
\end{figure}

We also investigated the polarization-dependence of 5-AGNR samples grown on Au(788) with a large number of short GNRs. As seen in the corresponding STM images in Figure~\ref{fig:FigSI_STM_LT-short5}, these ribbons are less well-aligned than GNRs grown with optimized parameters. This is reflected in the polarization anisotropy P, as shown in Figure~\ref{fig:FigSI_PolDep_short5}.

\clearpage

\newpage
\section{\large Supplementary Note 4~-~Damage sensitivity, substrate interaction and length distribution}\label{SI: Note 4 - Damage + Substrate}

\subsection*{\normalsize Radiation damaging of 9-AGNRs monitored \latin{via} the LCM}\label{SI: 4.2 Damaging}

Laser induced damage of nanomaterials is a well known problem in Raman spectroscopy. Above, we already discussed the mapping approach to mitigate this effect. 
On the other hand, the presence and shape of a Raman peak can be used as an indicator to assess the sample quality and investigate other sources of damage to the GNRs originating e.g. from sample fabrication.

In Figure~\sfigref[a]{fig:Fig4} we show the evolution of the peak intensity of the LCM and other modes commonly looked at. The underlying spectra are shown in Figure~\sfigref[a]{fig:FigSI_Damage_full} and are obtained by spatial averaging of subsequent scans of the same $\SI[product-units=power]{3 x 3}{\micro\meter}$-area with 5\,s integration time, each. Similarly, a power-dependent series (Figure~\sfigref[b]{fig:FigSI_Damage_full}) with averaging over a $\SI[product-units=power]{10 x 10}{\micro\meter}$-area was obtained. Integration time 1\,s at 0.5\,mW reduced to 0.1\,s at 40\,mW.

\newpage
\subsection*{\normalsize Substrate interaction}\label{SI: 4.3 Substrate-Interaction}

\begin{figure}[!htb]
    \centering
    \includegraphics[width=\textwidth]{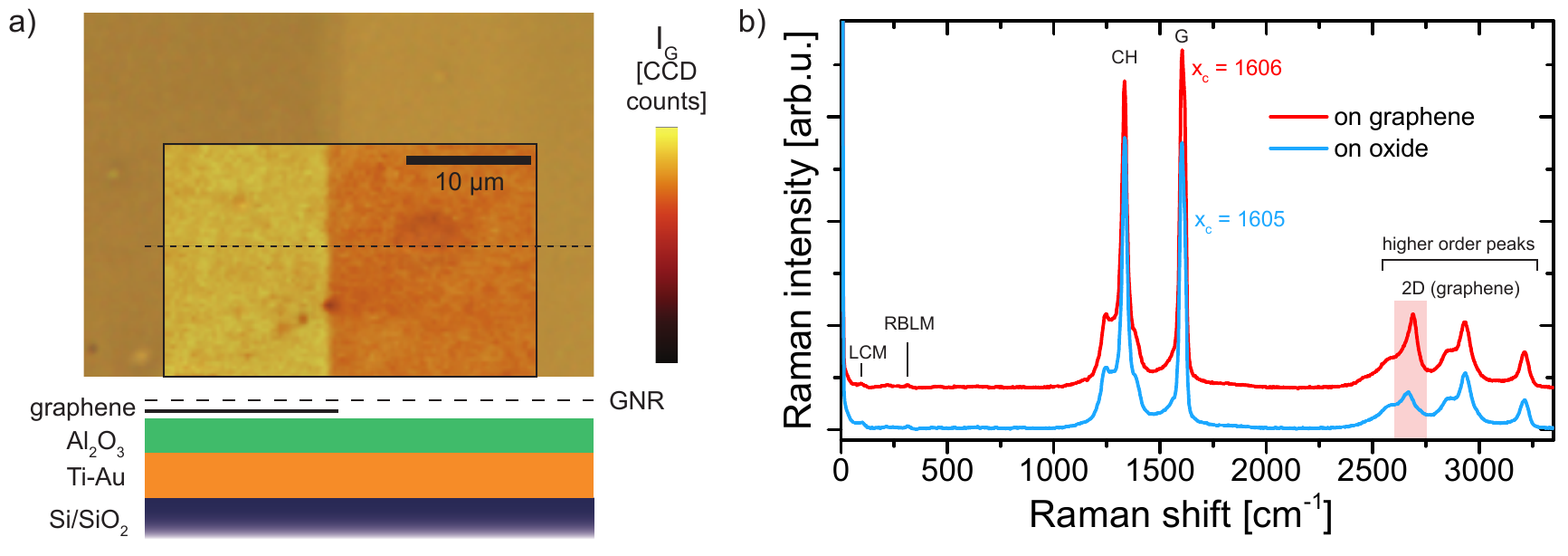} % Sample 9.I.06
    \caption{\textbf{Mapping to probe the effect of different substrates on the LCM. a)} Overlay of Raman map scan and optical microscope image showing GNRs on oxide (right) and on a graphene interlayer (left, darker contrast, higher G-peak intensity). Schematic of the sample layer structure. % map 11
    \textbf{b)} Full-range Raman spectrum extracted from a map scan as depicted in a and averaged over the respective substrate areas. G-peak position extracted from Lorentzian fit and graphene 2D-spectral region highlighted. P\tsub{exc}=1\,mW, $\lambda_{exc}$ = 488\,nm. Sample~9-1\_T.} % map 7
    \label{fig:FigSI_Substrate-Interaction}
\end{figure}

\begin{figure}[!htb]
\centering
\includegraphics[width=\textwidth]{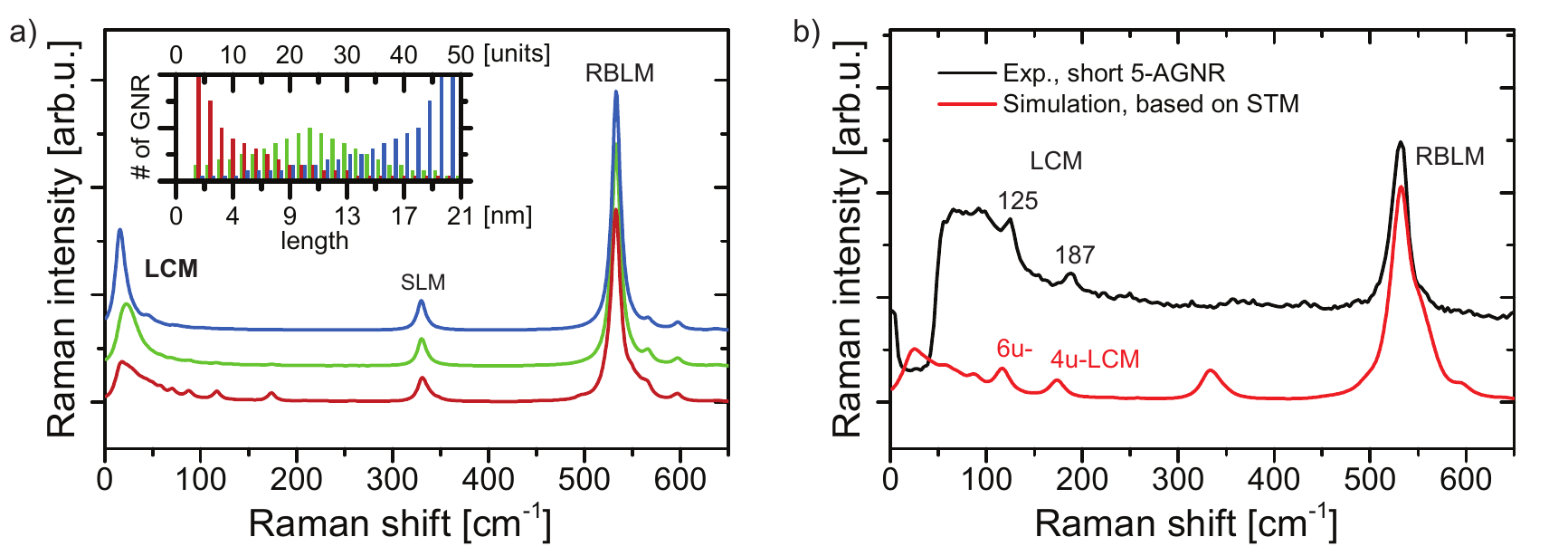}
\caption{\textbf{Simulated Raman spectra for 5-AGNR films. a)} Simulated Raman spectra of 5-AGNR films obtained by summing REBOII-based spectra for 5-AGNRs of different length, weighted according to the 3 length-distributions shown in the inset (colour coded short to long from bottom to top). b) Comparison of a simulated spectrum for a length distribution obtained from the STM of Sample~5-8\_Au in Figure~\sfigref[b]{fig:FigSI_STM_LT-short5} to the corresponding experimental Raman spectrum spectrum.}
\label{fig:FigSI_SummedMD}
\end{figure}

The substrate-dependent peak position of the LCM as shown in Figure~\ref{fig:Fig4} of the main manuscript was extracted by averaging GNR spectra acquired in a single map-scan covering both substrates, as shown in Figure~\ref{fig:FigSI_Substrate-Interaction}. An intensity map of the G-peak / 2D-peak area of graphene (stronger, additional signal compared to GNR-only) can be used to identify pixels with/without graphene interlayer. Note, that the G-peak shift is within our experimental uncertainty and tends in the opposite direction of the shift observed for the LCM.

\subsection*{\normalsize Summed REBOII Spectra}\label{SI: 4.1 Summing}

Typically, when experimental and calculated spectra are compared, the length distribution of ribbons is disregarded (as their aspect ratio is assumed to be very large).
As we argue here, this is an oversimplification and taking into account the length distribution of ribbons on a sample is key, in order to be able to match calculations with experiments.
As a first-order approximation, we can create simulated spectra for 5-AGNR films by taking the weighted sum of length-dependent REBOII-based spectra according to their percentage in the sample.
The result of this for three different (artificial) length distributions is shown in Figure \sfigref[a]{fig:FigSI_SummedMD}.

It can be seen, that this approach yields spectra with only minor variations for the RBLM, confirming that a length-independent treatment is valid for all samples with significant population of ribbons that are longer than 4 to 8 units, as only these shortest ribbons exhibit a splitting of the RBLM.
In contrast, the simulated spectrum shows that the LCM-region is sensitive to variations in the dominant fraction of ribbons above 10\,nm.
A simulated spectrum assuming a length distribution concentrated on very short ribbons as observed for the 5-AGNRs formed from partially dehydrogenated precursor molecules (see Figure~\sfigref[b]{fig:FigSI_STM_LT-short5} for corresponding STM image) is shown in panel b).
It qualitatively matches the experimental data obtained on these ribbons even though aspects of substrate interaction as well as resonance Raman effects have not been taken into account.
Improvements in these areas may in the future allow to extract the ribbon length distribution by fitting the weighting coefficients to experimental spectra.

 % If included here, referencing between manuscript and SI works both ways.

\end{document}